\documentclass[fleqn,usenatbib]{mnras}

\usepackage{newtxtext,newtxmath}

\usepackage[T1]{fontenc}
\usepackage{ae,aecompl}

\usepackage{graphicx}	
\usepackage{amsmath}	
\usepackage{amssymb}	

\usepackage{xcolor}
\usepackage{overpic}
\usepackage{subfigure}
\usepackage{booktabs}
\usepackage{multirow}
\usepackage{caption}
\usepackage[flushleft]{threeparttable}
\usepackage{natbib}

\newcommand{\bea}{\begin{eqnarray}}
\newcommand{\eea}{\end{eqnarray}}

\newcommand{\bra}[1]{\langle #1|}
\newcommand{\ket}[1]{|#1\rangle}
\newcommand{\braket}[2]{\langle #1|#2 \rangle}

\hypersetup{draft}

\title[Convective Excitation and Damping of Solar-like Oscillations]{Convective Excitation and Damping of Solar-like Oscillations}

\author[Zhou et al.]{
Yixiao Zhou,$^{1}$
Martin Asplund,$^{1}$
Remo Collet$^{2}$
and Meridith Joyce$^{1}$
\\
$^{1}$Research School of Astronomy and Astrophysics, Australian National University, Canberra, ACT 2611, Australia
\\
$^{2}$Stellar Astrophysics Centre, Department of Physics and Astronomy, Aarhus University, Ny Munkegade 120, DK-8000 Aarhus C, Denmark
\\
}

\date{Accepted XXX. Received YYY; in original form ZZZ}

\pubyear{2020}

\begin{document}
\label{firstpage}
\pagerange{\pageref{firstpage}--\pageref{lastpage}}
\maketitle

\begin{abstract}

The last decade has seen a rapid development in asteroseismology thanks to the \textit{CoRoT} and \textit{Kepler} missions. With more detailed asteroseismic observations available, it is becoming possible to infer exactly how oscillations are driven and dissipated in solar-type stars. We have carried out three-dimensional (3D) stellar atmosphere simulations together with one-dimensional (1D) stellar structural models of key benchmark turn-off and subgiant stars to study this problem from a theoretical perspective. Mode excitation and damping rates are extracted from 3D and 1D stellar models based on analytical expressions. Mode velocity amplitudes are determined by the balance between stochastic excitation and linear damping, which then allows the estimation of the frequency of maximum oscillation power, $\nu_{\max}$, for the first time based on \textit{ab initio} and parameter-free modelling. We have made detailed comparisons between our numerical results and observational data and achieved very encouraging agreement for all of our target stars. This opens the exciting prospect of using such realistic 3D hydrodynamical stellar models to predict solar-like oscillations across the HR-diagram, thereby enabling accurate estimates of stellar properties such as mass, radius and age.
  
\end{abstract}




\section{Introduction}

  Asteroseismology provides a unique window to revealing the physical properties of stars. For solar-like oscillations -- acoustic waves (so-called p-modes) excited and damped by near surface convection, global asteroseismic observables -- the large frequency separation $\Delta\nu$ and frequency of maximum oscillation power $\nu_{\max}$ are linked to the stellar radius and mass by the seismic scaling relations \citep{1991ApJ...368..599B,1995A&A...293...87K}. Measured individual oscillation frequencies can further constrain the physics of stellar interiors, for instance the size of the convection zone \citep{1997MNRAS.287..189B,2016A&A...589A..93D} or the rotation rate of the stellar core \citep{2012A&A...548A..10M}, which are difficult to probe by other means.
  In addition to global asteroseismic observables and individual oscillation frequencies, other observables such as mode amplitude and line width encrypt information about how oscillations are excited and damped in the star, a fundamental problem in stellar physics that is still not fully understood.

  Significant progress toward this problem has been made from the observational side thanks to high-quality asteroseismic data from the \textit{CoRoT} \citep{2008Sci...322..558M}, \textit{Kepler} \citep{2010Sci...327..977B} and \textit{TESS} \citep{2015JATIS...1a4003R} missions, as well as ground-based telescopes such as SONG (Stellar Oscillations Network Group, \citealt{2006MmSAI..77..458G}).
  With measured oscillation amplitudes and line widths available for thousands of solar-type oscillators, it is now possible to investigate the excitation and damping of p-mode oscillations across the Hertzsprung-Russell (HR) diagram. Indeed, empirical relations between oscillation amplitudes, mode line widths, and fundamental stellar parameters have been derived \citep{2009A&A...500L..21C,2011ApJ...743..143H,2012A&A...537A.134A,2017ApJ...835..172L,2018A&A...616A..94V} for main-sequence, subgiants and red giant stars. Oscillation amplitudes are proportional to the luminosity-mass ratio of the star, while line width, which is closely related to mode damping rate, increases with the effective temperature of the star. 
  
  On the theoretical side, realistic models of mode excitation and damping can illuminate the underlying physics of the oscillations. The first step is to model the mode excitation and damping for the Sun. Following the key insight by \citet{1977ApJ...212..243G} that solar p-mode oscillations are driven by turbulent convection, \citet{1992MNRAS.255..603B,1992MNRAS.255..639B} analytically quantified the excitation and damping rate of solar p-mode, then evaluated the oscillation amplitude for the Sun using the non-local mixing length theory (MLT) of convection in 1D models. His pioneering work demonstrated that solar five-minute oscillations are intrinsically stable and that the calculated mode damping rates and velocity amplitude agrees reasonably well with helioseismic observations. Based on \citet{1992MNRAS.255..603B}'s theoretical formulation, \citet{1999A&A...351..582H} studied excitation, damping rates and velocity amplitudes of solar-like oscillations for a grid of 1D models that corresponds to main-sequence stars. Their results indicate that the mode velocity amplitude is proportional to the luminosity-mass ratio for solar-type oscillators, quantitatively confirming the amplitude scaling relation proposed earlier by \citet{1995A&A...293...87K}. These and other (e.g.~\citealt{2001A&A...370..136S,2012A&A...540L...7B}) studies have improved our understanding of p-mode oscillations in solar-type stars. However, due to the lack of realistic theory of convection, which is ultimately driving the oscillations, the theoretical prescriptions adopted in these works inevitably involve adjustable parameters, making independent theoretical predictions of excitation and damping difficult.

  An alternative approach that has shown great promise for overcoming this difficulty is to evaluate excitation and damping rates from first principles using 3D hydrodynamical convection simulations. \citet{2001ApJ...546..576N} and \citet{2001ApJ...546..585S} extracted the excitation rates of solar radial modes directly from 3D simulations of near-surface layers of the solar convective region. \citet{2019ApJ...880...13Z} quantified both the excitation and damping rates for solar radial modes and estimated velocity amplitude and theoretical $\nu_{\max}$ for the Sun based on their 3D solar atmosphere model. Without introducing any tunable free parameters, very encouraging agreement between theoretical results and corresponding helioseismic observations are achieved with this approach. In this paper, we apply the theoretical formulation and numerical technique described in \citet{2019ApJ...880...13Z} to four other key benchmark turn-off and subgiant stars to examine whether our method is applicable to other solar-type oscillating stars or not and to investigate how excitation and damping processes vary across the HR diagram. 
  We also explore in detail how radial oscillations are damped in the near-surface region of stars, which has recently been studied by \citet{2019A&A...625A..20B} for the solar case.

\section{Modelling}

\begin{table*}
\begin{threeparttable}
\centering
\caption{Fundamental stellar parameters of KIC 6225718, Procyon, $\beta$ Hydri and $\delta$ Eri. Reference values are adopted from various literature, determined either from observation or detailed stellar modelling. The basic parameters of our 1D \textsc{mesa} models and 3D \textsc{Stagger} models for these stars are also shown. In 3D models, the effective temperature fluctuates over time therefore both mean effective temperature and its standard deviation are given.
\label{tb:param}}
{\begin{tabular*}{\textwidth}{@{\extracolsep{\fill}}ccccccc}
\toprule[2pt]

 \multicolumn{2}{c}{Stellar parameter} & $T_{\rm eff}$ [K] &  $\log g$ (cgs) &  [Fe/H] &  $M/M_{\odot}$ &  $R/R_{\odot}$ \\

\midrule[1pt]

\multirow{3}{*}{KIC 6225718}  & Reference & \text{$6230 \pm 60$ \hyperlink{ref:B2012}{(a)}}     & \text{$4.319^{+0.007}_{-0.005}$ \hyperlink{ref:L2017}{(b)}}             & \text{$-0.17 \pm 0.06$ \hyperlink{ref:B2012}{(a)}}      & \text{$1.10^{+0.04}_{-0.03}$ \hyperlink{ref:T2014}{(c)}}      & \text{$1.22 \pm 0.01$ \hyperlink{ref:T2014}{(c)}} \\
 
 & 1D model & 6217 & 4.318 & -0.14   & 1.14    & 1.225 \\
 
 & 3D model & $6231 \pm 14$ & 4.319 & 0   &      &  \\
 
\midrule[1pt]

\multirow{3}{*}{Procyon}  & Reference & \text{$6543 \pm 84$ \hyperlink{ref:A2005}{(d)}} & \text{$4.00 \pm 0.02$ \hyperlink{ref:H2015}{(e)}}             & \text{$-0.03 \pm 0.07$ \hyperlink{ref:BL2012}{(f)}} & \text{$1.461 \pm 0.025$ \hyperlink{ref:B2010}{(g)}}      &  \text{$2.031 \pm 0.013$ \hyperlink{ref:A2005}{(d)}} \\
 
 & 1D model & 6554 & 3.99 & 0.01   & 1.47     & 2.022 \\
 
 & 3D model & $6553 \pm 23$ & 4.00 & 0   &   &  \\
 
\midrule[1pt]

\multirow{3}{*}{$\beta$ Hydri}  & Reference & \text{$5873 \pm 45$ \hyperlink{ref:H2015}{(e)}} & \text{$3.98 \pm 0.02$ \hyperlink{ref:H2015}{(e)}} & \text{$-0.04 \pm 0.06$ \hyperlink{ref:H2015}{(e)}} & \text{1.04 \hyperlink{ref:B2011}{(h)}} & \text{$1.810 \pm 0.015$ \hyperlink{ref:B2010}{(g)}} \\

 & 1D model & 5861 & 3.96 & -0.06 & 1.09 & 1.814 \\
 
 & 3D model & $5893 \pm 14$ & 3.98 & 0 &  &  \\
 
\midrule[1pt]

\multirow{3}{*}{$\delta$ Eri}  & Reference & \text{$4954 \pm 30$ \hyperlink{ref:H2015}{(e)}} & \text{$3.76 \pm 0.02$ \hyperlink{ref:H2015}{(e)}} & \text{$0.06 \pm 0.05$ \hyperlink{ref:H2015}{(e)}} & \text{$1.13 \pm 0.05$ \hyperlink{ref:H2015}{(e)}} & \text{$2.327 \pm 0.029$ \hyperlink{ref:B2010}{(g)}} \\

 & 1D model & 4948 & 3.76 & 0.06 & 1.17 & 2.352 \\
 
 & 3D model & $4958 \pm 11$ & 3.76 & 0 &  &  \\
 
\bottomrule[2pt]
\end{tabular*}}

	\begin{tablenotes}
      \item Reference: \hypertarget{ref:B2012}{(a): \cite{2012MNRAS.423..122B}}; 
      \hypertarget{ref:L2017}{(b): \cite{2017ApJ...835..172L}}; 
      \hypertarget{ref:T2014}{(c): \cite{2014MNRAS.445.2999T}};
      \hypertarget{ref:A2005}{(d): \cite{2005ApJ...633..424A}};
      \hypertarget{ref:H2015}{(e): \cite{2015A&A...582A..49H}};
      \hypertarget{ref:BL2012}{(f): \cite{2012MNRAS.427...27B}};
      \hypertarget{ref:B2010}{(g): \cite{2010MNRAS.405.1907B}};
      \hypertarget{ref:B2011}{(h): \cite{2011A&A...527A..37B}}
    \end{tablenotes}
    
\end{threeparttable}
\end{table*}

\begin{table}
\begin{threeparttable}
\centering
\caption{Global asteroseismic parameters of KIC 6225718, Procyon, $\beta$ Hydri and $\delta$ Eri. Theoretical $\Delta\nu$ are derived following the method of \citet{2011ApJ...743..161W}. For Procyon, the value of $\nu_{\max}$ is uncertain because the observed oscillation spectrum exhibits a broad plateau between 600 and 1200 $\mu \rm Hz$ \citep{2008ApJ...687.1180A}.
\label{tb:seis_param}}
{\begin{tabular*}{1.05\columnwidth}{@{\extracolsep{\fill}}ccccc}
\toprule[2pt]
\multirow{2}{*}{Star} & \multicolumn{2}{c}{$\Delta\nu$ [$\mu \rm Hz$]} & \multicolumn{2}{c}{$\nu_{\max}$ [$\mu \rm Hz$]} \\

  & Observed & Modeling & Observed & Modeling$\dagger$ \\
\midrule[1pt]
  KIC 6225718 & 105.7 \hyperlink{ref:Lund17}{(a)} & 106.3 & 2364 \hyperlink{ref:Lund17}{(a)} & 2300
  \\ 
  Procyon & 55 \hyperlink{ref:Bedding10}{(b)} & 56 & --- & ---
  \\ 
  $\beta$ Hydri & 57.24 \hyperlink{ref:Bedding07}{(c)} & 58.37 & 1000 \hyperlink{ref:Bedding07}{(c)} & 980
  \\  
  $\delta$ Eri & 40.25 \hyperlink{ref:TESS}{(d)} & 40.45 & 677 \hyperlink{ref:TESS}{(d)}  & 650
  \\ 
\bottomrule[2pt]
\end{tabular*}}

	\begin{tablenotes}
      \item Reference: \hypertarget{ref:Lund17}{(a): \cite{2017ApJ...835..172L}}; 
      \hypertarget{ref:Bedding10}{(b): \cite{2010ApJ...713..935B}};
      \hypertarget{ref:Bedding07}{(c): \cite{2007ApJ...663.1315B}};
      \hypertarget{ref:TESS}{(d): TESS data (E.~Bellinger, in preparation)}
      
      $\dagger$: Note that the numbers listed here are only estimations, as  
      the exact value of theoretical $\nu_{\max}$ depends on how simulation data are 
      smoothed.
    \end{tablenotes}
    
\end{threeparttable}
\end{table}

\subsection{Target stars} \label{sec:star}

  The target stars investigated in this work are KIC 6225718, Procyon A, $\beta$ Hydri and $\delta$ Eridani ($\delta$ Eri). All of them are late-type, intermediate-mass (between $1M_{\odot}$ and $1.5M_{\odot}$) stars with metallicity [Fe/H]\footnote{${\rm [A/B]} = \log(n_{\rm A} / n_{\rm B}) - \log(n_{\rm A} / n_{\rm B})_{\odot}$ where $n_{\rm A} / n_{\rm B}$ and $(n_{\rm A} / n_{\rm B})_{\odot}$ represent number density ratio between element A and B in the star and the Sun, respectively.} near the solar value. Solar-like oscillations have been unambiguously detected for all of our targets, and well-determined $\Delta\nu$ and $\nu_{\max}$ are available. Their fundamental stellar parameters and global asteroseismic parameters are listed in Table \ref{tb:param} and \ref{tb:seis_param}, respectively. We introduce the basic properties of the four stars individually below.
  
  \

  \textbf{KIC 6225718 (HD 187637)} is an F-type main-sequence star observed by the \textit{Kepler} satellite for a long timespan. Based on high-quality \textit{Kepler} data, \citet{2012ApJ...757...99S} obtained global oscillation parameters, i.e.~$\Delta\nu$ and $\nu_{\max}$, for this star.
  More detailed studies on the oscillation properties of KIC 6225718 were carried out by \citet{2017ApJ...835..172L}, who identified more than 50 oscillation modes and determined their frequencies with uncertainties of typically $1 \; \rm\mu Hz$. Moreover, \citet{2017ApJ...835..172L} provided the measured line width and mode amplitude for each radial p-mode, which enables detailed comparison between observation and theoretical stellar models. Accurate atmospheric parameters of KIC 6225718 are also available from previous work. \citet{2012MNRAS.423..122B} determined the effective temperature of this star, $T_{\rm eff} = 6230 \pm 60$ K, and metallicity $\rm [Fe/H] = -0.17 \pm 0.06$ by fitting the stellar spectra with a fixed surface gravity ($\log g = 4.32$, cgs unit), which is determined from asteroseismology. 
  On the modelling side, KIC 6225718 has been investigated in detail by \citet{2014MNRAS.445.2999T}, \citet{2017ApJ...835..173S} and \citet{2019MNRAS.487..595H}. \citet{2014MNRAS.445.2999T} adopted the fundamental stellar parameters from \citet{2012MNRAS.423..122B} as basic constraints and measured $l=0,1,2$ p-mode frequencies as seismic constraints for their model. They estimated the most probable mass and radius of KIC 6225718 to be $M = 1.10_{-0.03}^{+0.04} M_{\odot}$, $R = 1.22 \pm 0.01 R_{\odot}$. \citet{2017ApJ...835..173S} modelled this star using observational constraints from \citet{2017ApJ...835..172L}. Their modelling involved various stellar evolution codes, input physics, fitting methods (see \citealt{2017ApJ...835..173S} Sect.~3 for details), resulting in a stellar mass $M = 1.2133 \pm 0.035 M_{\odot}$ and a radius of $R = 1.2543 \pm 0.0133 R_{\odot}$. \citet{2019MNRAS.487..595H}, on the other hand, focused on modelling the line width and corrections to adiabatic oscillation frequencies for KIC 6225718 using the non-local, time-dependent convection model \citep{1992MNRAS.255..603B,1999A&A...351..582H}.
  
  \

  \textbf{Procyon A (HD 61421)} is an F5 star with a white dwarf companion in a binary system and is one of the nearest stars to Earth. Owing to its proximity and brightness, Procyon A (hereinafter Procyon) is of particular importance to study.
  Therefore, much effort has been put into determining its fundamental parameters accurately. Among these efforts, \citet{2005ApJ...633..424A} measured the angular diameter of Procyon using interferometry. Together with the bolometric flux obtained from various instruments and the \textit{Hipparcos} parallax, they derived the radius and effective temperature to be $R = 2.031 \pm 0.013 R_{\odot}$ and $T_{\rm eff} = 6543 \pm 84$ K. The mass of Procyon is comparably well-constrained because it resides in a binary system. We adopt the orbital mass provided by \citet{2010MNRAS.405.1907B}: $M = 1.461 \pm 0.025 M_{\odot}$. The metallicity of Procyon has likewise been scrutinized in depth. \citet{2002ApJ...567..544A} analysed the spectrum of Procyon with a focus on the iron abundance determination from a 3D model atmosphere. Their 3D model was able to reproduce the observed Fe line profiles, yielding an iron abundance for Procyon of $\log\varepsilon_{\rm Fe} = 7.36 \pm 0.03$ dex, slightly lower than the solar value ($\log\varepsilon_{\rm Fe,\odot} = 7.41 \pm 0.02$ dex derived in the same paper). More recently, \citet{2012MNRAS.427...27B} performed 3D, non-local thermodynamic equilibrium (non-LTE) Fe line formation calculations for several late-type stars including Procyon based on up-to-date atomic data. With basic stellar parameters adopted from \citet{2005ApJ...633..424A}, they obtained the metallicity of Procyon to be ${\rm [Fe/H]} = -0.03 \pm 0.07$. Both works confirmed that Procyon is a solar-metallicity star.
  
  Procyon is also a favourable target for asteroseismology. Solar-like p-mode oscillations in Procyon were first announced by \citet{1991ApJ...368..599B}. More than a decade later, \citet{2008ApJ...687.1180A} conducted intensive radial velocity measurements for Procyon using 11 spectrographs at eight observatories. They found clear oscillation signatures from radial velocity variation and obtained velocity amplitude which demonstrated a plateau between 0.6 mHz and 1.2 mHz. \citet{2010ApJ...713..935B} subsequently extracted individual oscillation frequencies ranging from 0.3 mHz to 1.4 mHz, indicating a broad spectrum of stochastically excited p-modes in Procyon. Using individual mode frequencies as constraints, \cite{2010AN....331..949D} and \citet{2014ApJ...787..164G} performed asteroseismic modelling for Procyon. Both works predicted a stellar mass close to $1.5 M_{\odot}$. The latter also concluded that the star is still in the core-hydrogen burning phase.
  
  \

  \textbf{$\beta$ Hydri (HD 2151)} is the closest subgiant to Earth, making it an excellent object for investigation. 
  In this work we choose stellar parameters provided in \cite{2015A&A...582A..49H} as reference values: $T_{\rm eff} = 5873$ K, $\log g = 3.98$ dex, $\rm [Fe/H] = -0.04$ dex; The effective temperature is deduced from bolometric flux and angular diameter measured by \cite{2007MNRAS.380L..80N}. $\beta$ Hydri is a benchmark star in asteroseismology as well -- it is one of the first subgiants confirmed as a solar-type oscillator. \cite{2001ApJ...549L.105B} reported clear p-mode oscillation signatures in $\beta$ Hydri and estimated the frequency of maximum power around 1 mHz. The follow-up study by \cite{2007ApJ...663.1315B} further extracted individual mode frequencies and revealed the existence of mixed modes\footnote{In evolved stars such as subgiants and red giants, the evanescent layer between p-mode and g-mode cavity can be very thin, especially for $l=1$ modes. The coupling of oscillation cavities will result in a mixed character of some modes, which are excellent tools to probe the stellar core. See \cite{2017A&ARv..25....1H} for a thorough review.} in the star. \cite{2011A&A...527A..37B} subsequently reanalysed the observational data. With updated mode frequencies as constraints, they presented interior models for $\beta$ Hydri. 
  Additionally, we note that $\beta$ Hydri has recently been observed by TESS, which may supply more information about the oscillation properties of this star.
  
  \
 
  \textbf{$\delta$ Eri (HD 23249)} is a solar-metallicity K-subgiant included among the \textit{Gaia} benchmark stars \citep{2015A&A...582A..49H}. This star is ascending the red giant branch and its age is estimated to be $6-9$ Gyr \citep{2019MNRAS.482..895S}, which makes it both the most evolved and the oldest star in our sample. Solar-like oscillations of $\delta$ Eri were first reported by \cite{2003Ap&SS.284...21B} who found a clear oscillation signature around 0.7 mHz. It is worth noting that $\delta$ Eri has recently been observed simultaneously by SONG and TESS (E.~Bellinger, in preparation) so that both spectroscopic (radial velocity) and photometric (intensity) measurements of oscillation are available for this star, which allows detailed comparison between theoretical and measured oscillation properties.
  However, currently no measured mode line width data are available for $\delta$ Eri. Therefore, theoretical damping rates of this star are compared with measured line widths of KIC 5689820 (cf.~Sect.~\ref{sec:mode} and \ref{sec:compobs}), a subgiant observed by \textit{Kepler}. The basic stellar parameters of KIC 5689820 are $T_{\rm eff} = 5037 \pm 76$ K, $\log g = 3.76 \pm 0.06$ dex, $\rm [Fe/H] = 0.21 \pm 0.15$ dex \citep{2020arXiv200506460L}, suggesting it is analogous to $\delta$ Eri. As such, its oscillation properties are likely to be comparable to our simulation results.

\subsection{Three-dimensional stellar atmosphere models} \label{sec:3D}

\begin{table*}
\begin{threeparttable}
\centering
\caption{Basic information about the set-up of 3D simulations for KIC 6225718, Procyon, $\beta$ Hydri and $\delta$ Eri. ``Sampling interval'' refers to the time interval between two consecutive simulation snapshots. As mesh points are not uniformly distributed vertically, both minimum and maximum vertical grid spacing are provided.
\label{tb:3Dsetup}}
{\begin{tabular*}{\textwidth}{@{\extracolsep{\fill}}ccccccccc}
\toprule[2pt]
  \multirow{3}{*}{Model configuration} & \multicolumn{2}{c}{KIC 6225718} & \multicolumn{2}{c}{Procyon} & \multicolumn{2}{c}{$\beta$ Hydri} & \multicolumn{2}{c}{$\delta$ Eri}
  \\ 
   & \multirow{2}{*}{Normal} & Artificial & \multirow{2}{*}{Normal} & Artificial & \multirow{2}{*}{Normal} & Artificial & \multirow{2}{*}{Normal} & Artificial
  \\
   & & driving & & driving & & driving & & driving 
  \\
\midrule[1pt]
  Resolution  & $240^3$ & $120^2 \times 125$ & $240^3$ & $120^2 \times 125$ & $240^3$ & $120^2 \times 125$ & $240^3$ & $240^3$
  \\
  Time duration $\mathcal{T}_{\rm tot}$ [hour] & 30.0 & 9.0 & 43.5 & 29.0 & 58.3 & 17.5 & 77.5 & 31.0 
  \\
  $\mathcal{T}_{\rm tot} / t_{\rm gran,eff}$ $^{\dagger}$ & 261 & 78 & 207 & 138 & 255 & 77 & 206 & 82 
  \\
  Sampling interval [s] & 36 & 36 & 87 & 87 & 70 & 70 & 93 & 93
  \\
  Vertical size [Mm] & 5.9 & 5.9 & 17.5 & 17.5 & 11.6 & 11.6 & 14.4 & 14.4
  \\
  Vertical grid spacing [km]   & 15--51 & 31--103 & 31--244 & 63--489 & 29--93 
  & 57--185 & 43--109 & 43--109
  \\
  Horizontal grid spacing [km] & 53     & 106     & 126     & 251     & 110    
  & 221     & 142     & 142
  \\
\bottomrule[2pt]
\end{tabular*}}

	\begin{tablenotes}
      \item $\dagger$: The granulation timescale $t_{\rm gran,eff}$ is the $e$-folding time of a granule. It is estimated from the empirical relation $t_{\rm gran,eff} = 2 \times 10^6 g^{-0.85} (T_{\rm eff} / 5777)^{-0.4}$ ($T_{\rm eff}$ and $g$ in cgs unit) which is calibrated from a wide variety of stars observed by \textit{Kepler}. See \citet{2014A&A...570A..41K} for more detail.
    \end{tablenotes}
    
\end{threeparttable}
\end{table*}

  In this section, we briefly describe the 3D hydrodynamical model atmospheres constructed for the target stars. All 3D models are computed with a customized version of the \textsc{Stagger} code \citep{1995...Staggercodepaper,2018MNRAS.475.3369C}, a radiative-magnetohydrodynamics code that solves the equations of mass, momentum, and energy conservation, as well as the magnetic-field induction equation on 3D Eulerian meshes. All scalars are evaluated at cell centres while vectors are staggered at the cell faces. The radiative heating rate in the equation of energy conservation is obtained by solving the 3D equation of radiative transfer along a set of inclined rays in space, assuming LTE. 
  In total, nine directions -- one vertical direction and eight inclined directions representing combinations of two polar and four azimuthal angles -- are included for all models presented in this work.
   The code is equipped with realistic microphysics: it uses a modified version of the \citet{1988ApJ...331..815M} equation of state \citep{2013ApJ...769...18T} that accounts for all ionization stages of the 17 most abundant elements in the Sun as well as the Hydrogen molecule. A comprehensive collection of relevant continuous absorption and scattering is included \citep{2010A&A...517A..49H}. Line opacities are taken from the MARCS model atmosphere package \citep{2008A&A...486..951G} and treated with the opacity binning method \citep{1982A&A...107....1N,2013A&A...557A..26M}, with 12 opacity bins divided based on both wavelength and strength of opacity.

  Our \textsc{Stagger} model stellar atmosphere simulates a small part of the star near the photosphere, assuming a constant gravitational acceleration and ignoring magnetic field. Geometrically, the simulation domain is discretized on a cuboid box. The horizontal size of the box is required to be large enough to enclose at least ten granules at any time in the simulation \citep{2013A&A...557A..26M}.
  Vertically, the 3D simulation covers roughly the outer 1\% of the star by radius, where hydrodynamical and 3D effects are most prominent. Because the vertical (radial) scale of the simulation is very small compared to the total stellar radius, the spherical effect in simulation domain is negligible. Boundaries are periodic in the horizontal direction while open in the vertical \citep{2018MNRAS.475.3369C}. The default bottom boundary condition is that outgoing flows (vertical velocities towards stellar centre) are free to carry their entropy fluctuations out of the simulation domain, whereas incoming flows (vertical velocities towards stellar surface) must have invariant entropy and thermal (gas plus radiation) pressure. The 3D model\footnote{When stating ``3D simulations/models'' or ``normal simulations/models'', we always mean the simulation carried out with the default boundary condition in the \textsc{Stagger} code.} for each star is constructed based on the reference $T_{\rm eff}$ and $\log g$ values given in Table \ref{tb:param}, and their basic properties are summarised in Table \ref{tb:3Dsetup}. All models adopt the \citet{2009ARA&A..47..481A} solar abundance, as all our targets are solar-metallicity stars. 
  The spatial resolution of our models is $240^3$, with $240 \times 240$ mesh points evenly distributed in the horizontal plane. In the vertical direction, mesh points are not uniformly distributed. The highest numerical resolution is applied around the photosphere (\citealt{2013A&A...557A..26M} Fig.~2) in order to resolve the transition from the optically thick to the optically thin regions adequately, with at least 15 mesh points per pressure scale height around the photosphere in all 3D models.
  The adopted spatial resolution should be sufficient to study the mode excitation problem, as differences between excitation rates computed from $253 \times 253 \times 163$ and $125 \times 125 \times 82$ solar simulations are small, according to \citet{2007A&A...463..297S}. The duration of the simulation is long enough to cover at least 200 times the granulation timescale. The mean effective temperature over the entire simulation timespan for each 3D model is close to the reference value.

  In addition, as mentioned in \citet{1998IAUS..185..199N} and \citet{2019ApJ...880...13Z}, it is non-trivial to extract reliable damping rates from 3D simulations. To overcome this difficulty, we conduct numerical experiments that artificially drive radial oscillation at a particular frequency to large amplitude by modifying the bottom boundary condition. The artificial driving simulation enables reliable calculation of damping rate at the driving frequency. For each star, we carried out such experiments at various driving frequencies in order to obtain damping rates as a function of frequency. For the purpose of controlling variables, all artificial driving experiments for a given star share the same input options with driving frequency being the only difference, and their numerical resolution and time sequence are also identical. Note that in the case of $\delta$ Eri, we adopt the standard $240^3$ resolution, whereas for the three other stars, we reduce the numerical resolution to $120^2 \times 125$ (120 by 120 cells in the horizontal plane with 125 points along vertical direction). The underlying reasons and validation for lowering the resolution are discussed in Appendix \ref{sec:etanum}, where we also detail numerical techniques about the artificial mode driving simulation, including tests and validation of our method.

\subsection{One-dimensional stellar interior models} \label{sec:1D}

\begin{table*}
	\centering
	\caption{Summary of fitting targets and free parameters in stellar interior models.}
	\label{tb:fit_params}
	\begin{tabular}{ccc} 
    \toprule[2pt]
Input free parameters                           & Targets to fit 
\\
	\midrule[1pt]
Stellar mass $M$ & Effective temperature $T_{\rm eff}$
\\
Initial Metallicity $\rm [Fe/H]_{init}$         & Surface gravity $\log g$            
\\
Initial Helium mass fraction $Y_{\rm init}$     & Metallicity $\rm [Fe/H]$    
\\
Mixing length parameter multiplier $f_{\alpha}$ & Averaged 3D temperature at matching point $\langle \bar{T}_{\rm 3D}(r_{\rm am}) \rangle_t$  
\\             
Convective turbulence multiplier $\beta$        & Averaged 3D turbulent pressure at matching point $\langle \bar{P}_{\rm turb,3D}(r_{\rm am}) \rangle_t$
\\
Convective overshoot parameter $f_{\rm ov}$     & 
\\
	\bottomrule[2pt]
	\end{tabular}
\end{table*}

\begin{figure}
\begin{overpic}[width=\columnwidth]{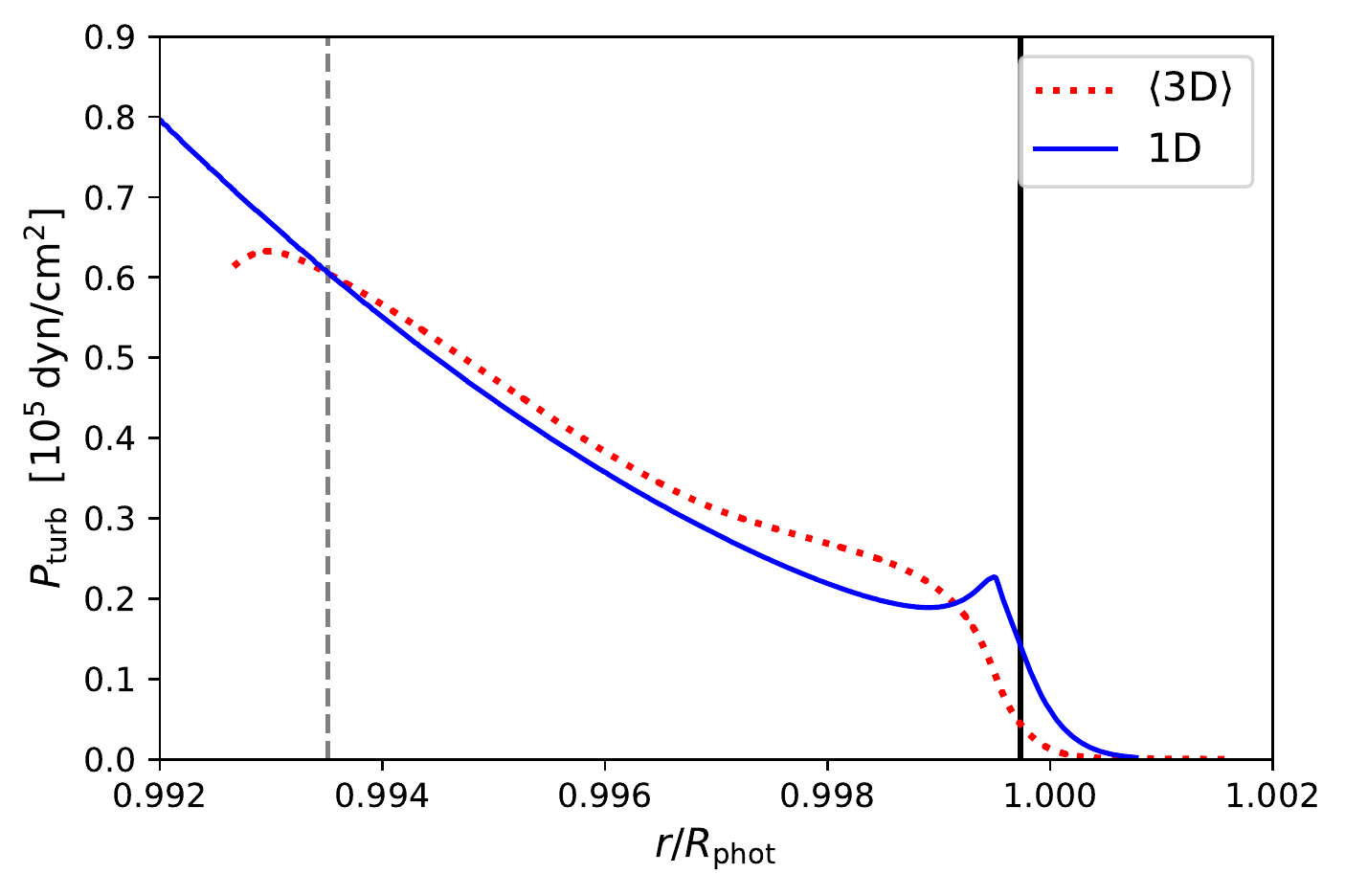}
\end{overpic}
\caption{Predicted distribution of turbulent pressure in the near-surface region of $\delta$ Eri models. The blue solid line represents 1D ``turbulent pressure'' calculated from \textsc{mesa} with modifications introduced in Sect.~\ref{sec:1D} while the red dotted line is horizontal- and time-averaged turbulent pressure from 3D atmosphere model of $\delta$ Eri. Grey dashed vertical line indicates the location of the matching point, and black solid vertical line marks the upper convection boundary $r_{\rm edge}$ in the \textsc{mesa} model. $R_{\rm phot}$ is the stellar radius.}
\label{fig:match_Pturb}
\end{figure}

  The 1D interior models for the target stars are computed using the Modules for Experiments in Stellar Astrophysics (\textsc{mesa} version 10000, \citealt{2011ApJS..192....3P,2013ApJS..208....4P,2015ApJS..220...15P}). For all calculations, we adopt the \citet{2009ARA&A..47..481A} metal mixture. Equation of state tables are generated from the FreeEOS\footnote{\url{http://freeeos.sourceforge.net/}} code \citep{2012ascl.soft11002I}, which takes into account 20 elements (Hydrogen, Helium and 18 metals) and all of their ionization stages in its calculations. At low temperatures ($\log T < 4.5$), continuous and line opacities at each wavelength are calculated from \textsc{blue}, an opacity package that adopts up-to-date atomic data (developed primarily for detailed non-LTE radiative transfer calculations, see \citealt{2016MNRAS.463.1518A} for description of the code). These are then reduced to Rosseland mean opacities for use in \textsc{mesa}. Opacities at high temperatures ($\log T > 4.4$) are taken from the OPAL tables \citep{1996ApJ...464..943I}. The two sets of opacities are blended in the temperature interval $4.4 < \log T < 4.5$. Nuclear reaction rates are from JINA REACLIB \citep{2010ApJS..189..240C} plus additional tabulated weak reaction rates \citep{1985ApJ...293....1F, 1994ADNDT..56..231O, 2000NuPhA.673..481L}, which is the default setting in \textsc{mesa}. 
Element diffusion and gravitational settling are treated following the default method in \textsc{mesa} (see \citealt{2011ApJS..192....3P} Sect.~5.4). However, for stars more massive than $\approx 1.4 M_{\odot}$, considering element diffusion and gravitational settling alone will result in near or complete depletion of helium and heavy elements at the stellar surface in some stages of their evolution (see e.g., Fig.~1 of \citealt{2019MNRAS.489.1850V}), which contradicts measured abundances of F-type stars in open clusters \citep{1999A&A...351..247V}. 
To counter the effects of element diffusion and gravitational settling below the surface convection zone, we include turbulent diffusion in our calculations. The turbulent diffusion coefficient is computed according to \cite{2017ApJ...840...99D} for every time-step during the evolution calculations.

In \textsc{mesa}, thermal (gas plus radiation) pressure $P_{\rm ther}$ and temperature $T$ at the outermost cell (surface) are required for the outer boundary conditions (\citealt{2011ApJS..192....3P} Sect.~5.3). Here, we place the outer boundary of the \textsc{mesa} models above the photosphere (the Rosseland mean optical depth at surface $\tau_{\rm surf} \approx 5\times 10^{-3}$). Instead of integrating the Eddington grey atmosphere and using the Eddington $T - \tau$ relation to obtain pressure and temperature at the outer boundary, we opt for $P_{\rm ther}$ and $T$ derived from 3D simulations. Specifically, the $P_{\rm ther} - \tau$ and $T - \tau$ relations are computed from the \textsc{Stagger}-grid \citep{2013A&A...557A..26M} that spans a wide range of stellar parameters. For each \textsc{Stagger}-grid model, we extract $P_{\rm ther} - \tau$ relation by horizontally and temporally averaging the 3D thermal pressure and (Rosseland mean) optical depth. Temperature stratifications are calculated following the method developed in \cite{2014MNRAS.442..805T}, which gives the predicted $T - \tau$ relation when all heat is transported by radiation. Following the aforementioned procedure, we obtain $P_{\rm ther} - \tau$ and $T - \tau$ relations at various $\left\{ T_{\rm eff}, \; \log g, \; \rm [Fe/H] \right\}$ combinations. The results are then tabulated and applied in the evolutionary simulation: At every iteration during the model's evolution, the $P_{\rm ther} - \tau$ and $T - \tau$ relations corresponding to the current stellar parameters are computed by interpolation. Surface pressure and temperature are subsequently evaluated at the given $\tau_{\rm surf}$.
  
  \textsc{mesa} adopts the diffusion approximation for radiative transfer, which is valid in the stellar interior but is not a satisfactory approximation when $\tau \lesssim 10$ \citep{2014MNRAS.442..805T}. As the outer boundary of our stellar model is located above the photosphere, we correct the radiative temperature gradient (down to $\tau = 10$) to obtain a more realistic temperature structure in the atmosphere portion of the \textsc{mesa} model. The corrected radiative temperature gradient reads \citep{2014MNRAS.442..805T,2018MNRAS.478.5650M}:
\begin{equation}
\nabla_{\rm rad} = \nabla_{\rm rad, \textsc{mesa}} \left[ \frac{dq(\tau)}{d\tau} + 1 \right],
\end{equation}
with $q$ as the Hopf function: 
\begin{equation}
q(\tau) = \frac{4}{3} \left[ \frac{T(\tau)}{T_{\rm eff}} \right]^4 - \tau.
\end{equation}
The term $\nabla_{\rm rad, \textsc{mesa}}$ is the original \textsc{mesa} radiative temperature gradient computed based on the diffusion approximation, and $T(\tau)$ represents the $T - \tau$ relation when all heat is transported by radiation.

  Convection is treated using the \cite{1965ApJ...142..841H} formulation of the MLT. We do not treat the mixing length parameter $\alpha_{\rm MLT}$ as a constant throughout the star's evolution, but rather as a varying quantity which depends on effective temperature, surface gravity and metallicity (see also \citealt{2018MNRAS.478.5650M} Eq.~2):
\begin{equation}
\alpha_{\rm MLT}\left( T_{\rm eff}, \log g, {\rm [Fe/H]} \right) = 
f_{\alpha} \alpha_{\rm MLT,3D}\left( T_{\rm eff}, \log g, {\rm [Fe/H]} \right).
\end{equation} 
Here $\alpha_{\rm MLT,3D}$ is the mixing length parameter calibrated from the \textsc{Stagger}-grid (value taken from \citealt{2015A&A...573A..89M}), interpolated to current $T_{\rm eff}$, $\log g$ and $\rm [Fe/H]$. By including the free parameter $f_{\alpha}$, the mixing length parameter multiplier, we ensure that the relative value of $\alpha_{\rm MLT}$ in \textsc{mesa} is consistent with the results calibrated from the \textsc{Stagger}-grid, but allow its absolute value to differ in order to account for the different equations of state and opacities between the \textsc{Stagger}-code and \textsc{mesa}. 
In other words, we retain the variation of $\alpha_{\rm MLT}$ across the HR diagram as indicated by 3D surface convection simulations but with an absolute value consistent with the solar calibration using the aforementioned equations of state and opacities.

  Further, we include the ``turbulent pressure'' term, which is typically ignored in 1D hydrostatic models in stellar evolution calculations (see \citealt{2014MNRAS.445.4366T} and \citealt{2019MNRAS.488.3463J} for efforts in this direction). The 1D ``turbulent pressure'' is constructed based on the convective velocity $v_{\rm conv}$ from MLT:
\begin{equation}
P_{\rm turb,1D}(r) = \beta \rho(r)v_{\rm conv}^2(r),
\end{equation}
where $r$ denotes radius, $\rho$ is mass density. The convective turbulence multiplier $\beta$ is a free parameter to be specified before the evolutionary calculations. The value of $\beta$ is determined by requiring that the ``turbulent pressure'' in \textsc{mesa} and the horizontal- and time-averaged 3D turbulent pressure are identical at the matching point (detailed below). However, $v_{\rm conv}$ predicted from MLT will rapidly decrease to zero when approaching the convection boundary, which causes a sudden drop of $P_{\rm turb,1D}$ (see e.g., Fig.~3 of \citealt{2014MNRAS.445.4366T}) and therefore an unrealistically large turbulent pressure gradient. In order to overcome this problem, we consider the influence of convective overshoot on $v_{\rm conv}$. 
Overshoot becomes relevant at a location near the convection boundary in the convection zone and extends the convective velocity beyond the top convection boundary (the overshoot region) where an exponential decay of $v_{\rm conv}$ is assumed (\citealt{2011ApJS..192....3P} Eq.~2),
\begin{equation}
v_{\rm conv}(r) = v_{\rm conv}(r_0) 
\exp\left[-\frac{2|r-r_0|}{f_{\rm ov}H_P(r_{\rm edge})}\right].
\end{equation}
Here, $r_0$ is the location where overshoot starts to take effect, and $r_{\rm edge}$ is the corresponding radius of upper convection boundary. $H_P(r_{\rm edge})$ is the pressure scale height at upper convection boundary, and $f_{\rm ov}$ is called overshoot parameter. We calibrate $f_{\rm ov}$ at the upper boundary of the surface convection zone using the horizontal- and time-averaged 3D turbulent pressure from the \textsc{Stagger}-grid\footnote{At all other convection boundaries, such as the bottom boundary of surface convection zone, $f_{\rm ov}$ is still a free parameter.}. 
As demonstrated in Fig.~\ref{fig:match_Pturb}, the inclusion of convective overshoot and a suitable value of the overshoot parameter (at the top of surface convection zone) ensure a gradual change of $P_{\rm turb,1D}$ near the convection boundary, thus bringing the 1D ``turbulent pressure'' profile into better agreement with the averaged turbulent pressure predicted from 3D simulations.

  We carried out evolutionary calculations from the pre-main-sequence to the age at which target stellar parameters are satisfied assuming the aforementioned input physics. To obtain a reliable stellar interior model, we minimize the difference between model parameters and corresponding constraints by iteratively adjusting the free parameters in \textsc{mesa}. Basic stellar parameters of our best-fitting models are presented in Table \ref{tb:param} for the four target stars.
Input free parameters and constraints are listed in Table \ref{tb:fit_params}. The parameters $\langle \bar{T}_{\rm 3D}(r_{\rm am}) \rangle_t$ and $\langle \bar{P}_{\rm turb, 3D}(r_{\rm am}) \rangle_t$ (the bar symbol and $\langle ... \rangle_t$ represent the horizontal average and temporal average, respectively) are two extra constraints from the 3D models, where $r_{\rm am}$ symbolizes the location of matching point in the 3D model. The matching point in \textsc{mesa}, $r_{\rm im}$, is determined by requiring identical thermal pressures between the 1D and averaged 3D models,
\begin{equation}
P_{\rm ther,1D}(r_{\rm im}) = \langle \bar{P}_{\rm ther, 3D}(r_{\rm am}) \rangle_{t}.
\end{equation}
Fitting the averaged 3D temperature and turbulent pressure not only tightly restricts $f_{\alpha}$ and $\beta$, but it is also necessary for smooth transitions (for temperature and total pressure) from the interior model to the atmosphere model. This is essential for the patching procedure described in the subsequent section.

\subsection{Oscillation frequencies from patched models}

\begin{figure*}
\begin{overpic}[width=0.9\textwidth]{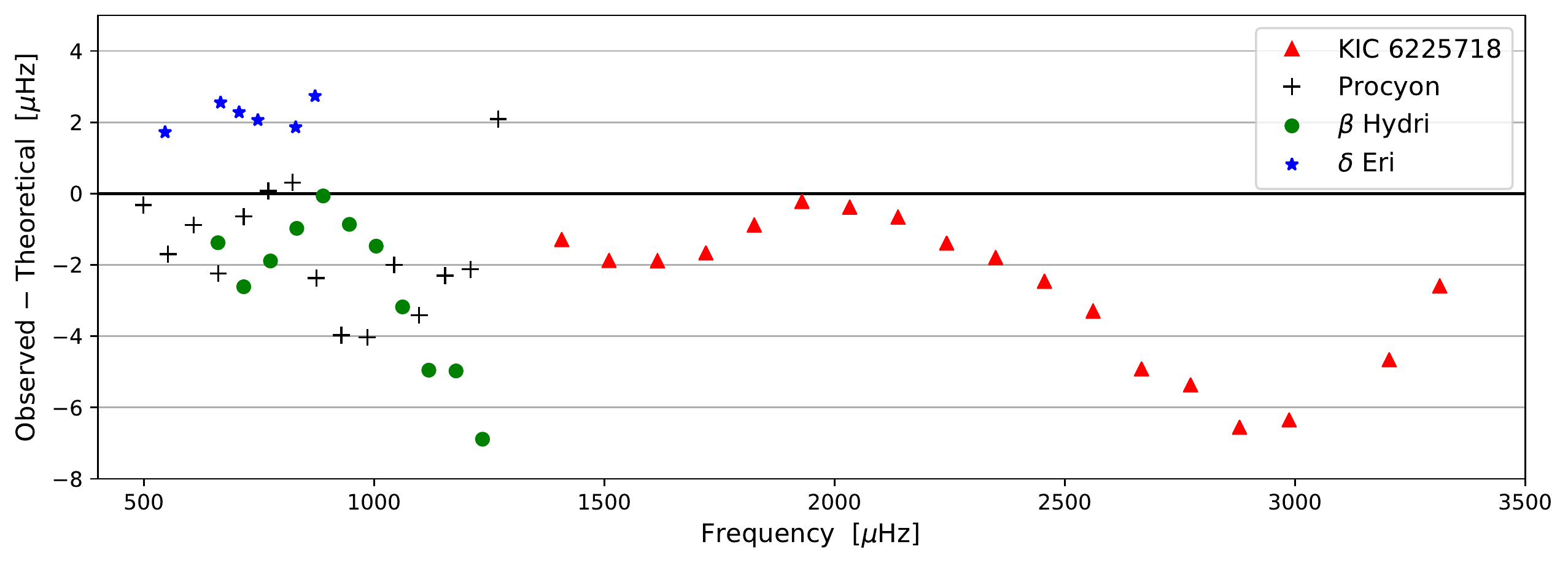}
\end{overpic}
\caption{Frequency differences between observations and theoretical results computed with patched $\rm 3D + 1D$ models. Only radial ($l=0$) modes are compared here. Observational data for KIC 6225718, Procyon, $\beta$ Hydri and $\delta$ Eri are from \citet{2017ApJ...835..172L}, \citet[table 4]{2010ApJ...713..935B}, \citet{2007ApJ...663.1315B} and E.~Bellinger (private communication), respectively.}
\label{fig:freq_diff}
\end{figure*}

  We combine the horizontally and temporally averaged 3D models and 1D interior models to patched 1D models for a more precise calculation of the oscillation properties (i.e.~eigenfrequencies, eigenfunctions, mode masses etc.). The fitting method introduced in Sect.~\ref{sec:1D} ensures continuous transitions in temperature and total pressure between $r_{\rm am}$ and $r_{\rm im}$, hence make patching straightforward in practice: The averaged 3D model and best 1D model for the same star are trimmed by discarding all layers below the atmosphere matching point in the averaged 3D model, and all layers above the interior matching point in the 1D model. They are then conjoined to obtain the patched 1D model that ranges from the stellar centre to the upper atmosphere ($\tau \sim 10^{-6}$). Pulsation calculations are performed with the Aarhus adiabatic oscillation package (\textsc{adipls}, \citealt{2008Ap&SS.316..113C}) using the patched model as input. Theoretical radial mode frequencies are compared with measured values for all stars investigated, and reasonable agreement is found between the two in each case as demonstrated in Fig.~\ref{fig:freq_diff}. The agreement in individual (radial) mode frequencies, in conjunction with the fact that $T_{\rm eff}$, $\log g$ and $\rm [Fe/H]$ of 1D model are close to the observationally inferred values, indicates that our patched models realistically describes the structure of target stars.

\section{Mode excitation, damping and amplitude} \label{sec:mode}


\begin{figure*}
\begin{overpic}[width=0.9\textwidth]{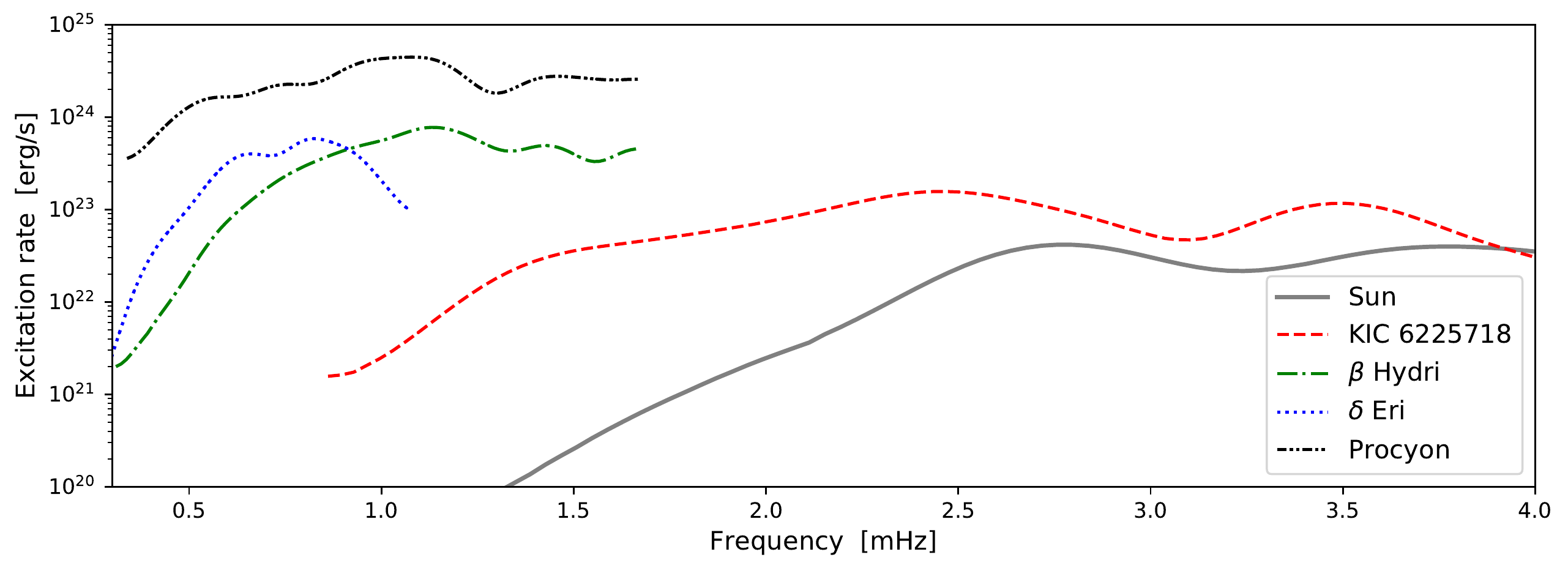}
\end{overpic}
\caption{Excitation rates as a function of cyclic frequency for our target stars computed via Eq.~\eqref{eq:er_num}. Theoretical excitation rate of the Sun (see Fig.~6 of \citealt{2019ApJ...880...13Z}) is also shown for comparison. The curves are smoothed from the original simulation data with Gaussian kernels whose FWHM are 0.47, 0.39, 0.16, 0.16, 0.11 mHz for the Sun, KIC 6225718, Procyon, $\beta$ Hydri, $\delta$ Eri, respectively.}
\label{fig:er}
\end{figure*}


\begin{figure}
\begin{overpic}[width=\columnwidth]{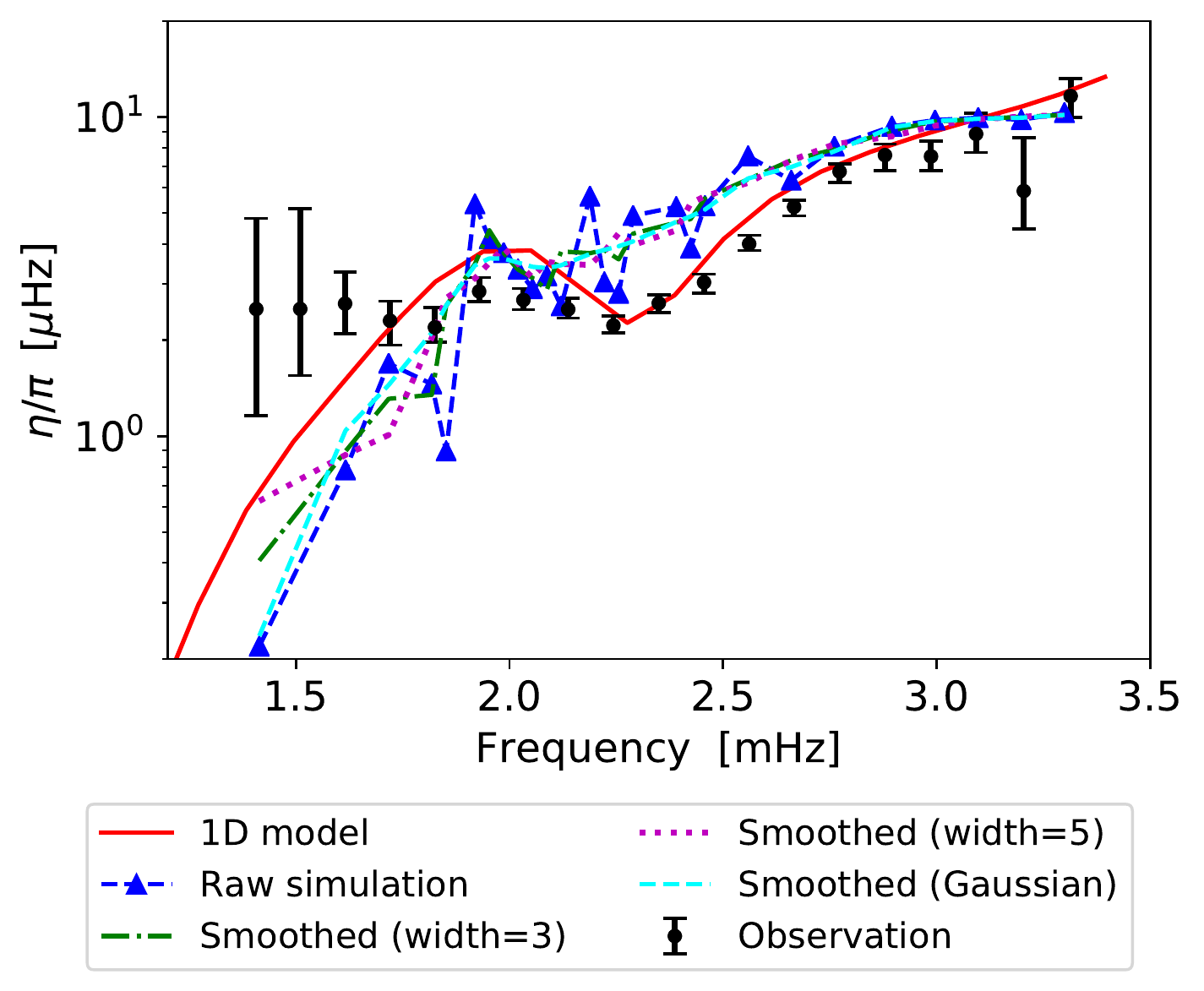}
\end{overpic}
\caption{Theoretical damping rates for KIC 6225718 computed based on Eq.~\eqref{eq:etamain}. Simulation results (raw data are blue triangles, smoothed data are green dashed-dotted, magenta dotted, and cyan dashed lines) are divided by $\pi$ to compare with observed $l=0$ mode line width (black dots with errorbars representing uncertainty) from \citet{2017ApJ...835..172L}. The green and magenta curves are obtained by taking the running mean of the raw simulation data with a width of three and five data points, respectively, whereas the cyan dashed line results from smoothing the raw simulation data by a Gaussian kernel with an FWHM equal to 0.18 mHz. The red solid line represents $\eta / \pi$ for this star computed from 1D non-local, time-dependent convection model \citep{2019MNRAS.487..595H}.}
\label{fig:t62g43m00eta}
\end{figure}

\begin{figure}
\begin{overpic}[width=\columnwidth]{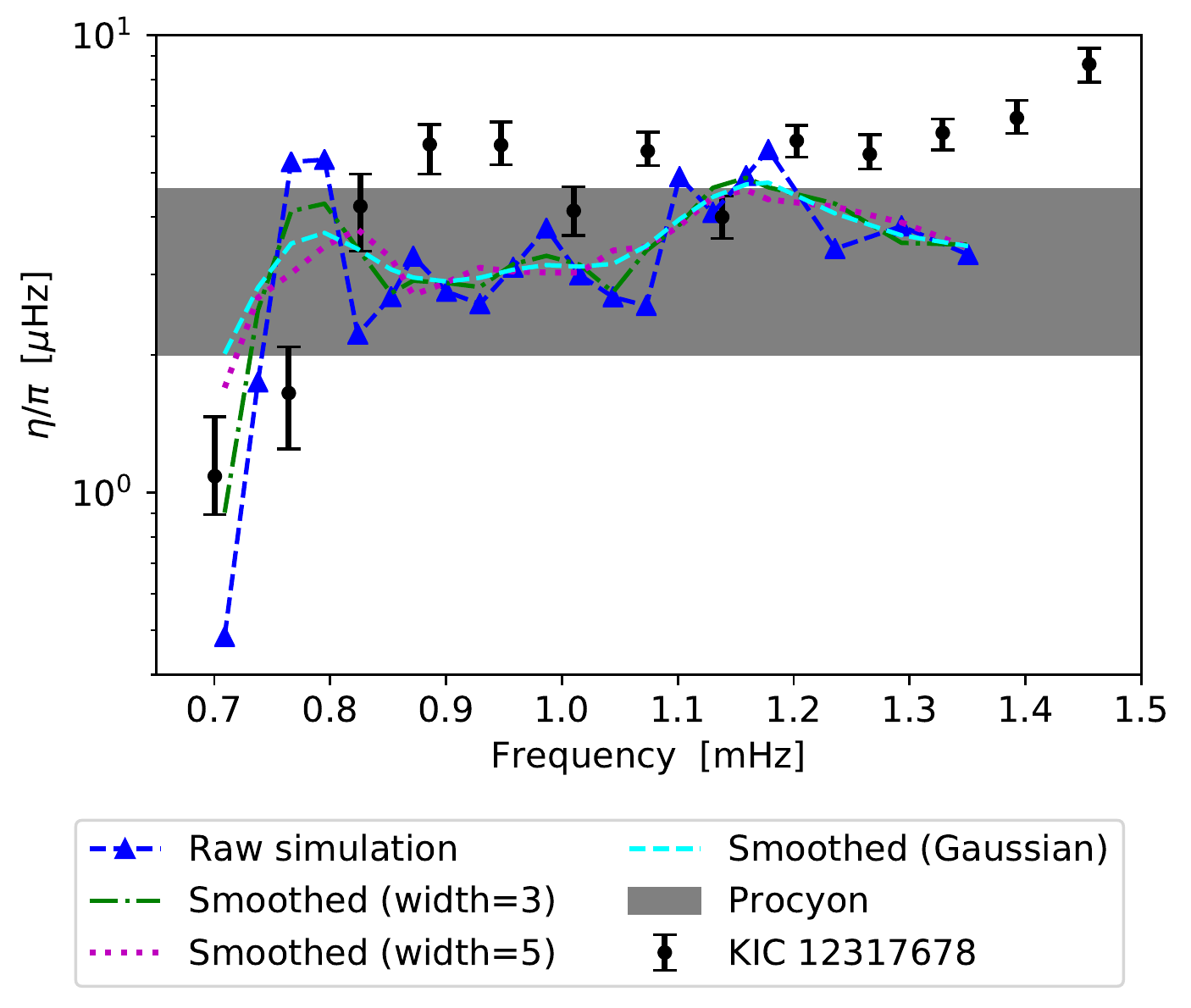}
\end{overpic}
\caption{Theoretical damping rates for Procyon. The cyan dashed line is smoothed from raw simulation data by a Gaussian kernel with an FWHM equal to 0.1 mHz. The grey-shaded band is the measured mean line width (with uncertainty) converted from the mean mode lifetime provided in \citet{2010ApJ...713..935B}. Black dots and errorbars are radial mode line width of star KIC 12317678 measured by \citet{2017ApJ...835..172L}. The fundamental parameters of KIC 12317678 are close to Procyon therefore it is shown for comparison as well.}
\label{fig:t66g40m00eta}
\end{figure}

\begin{figure}
\begin{overpic}[width=\columnwidth]{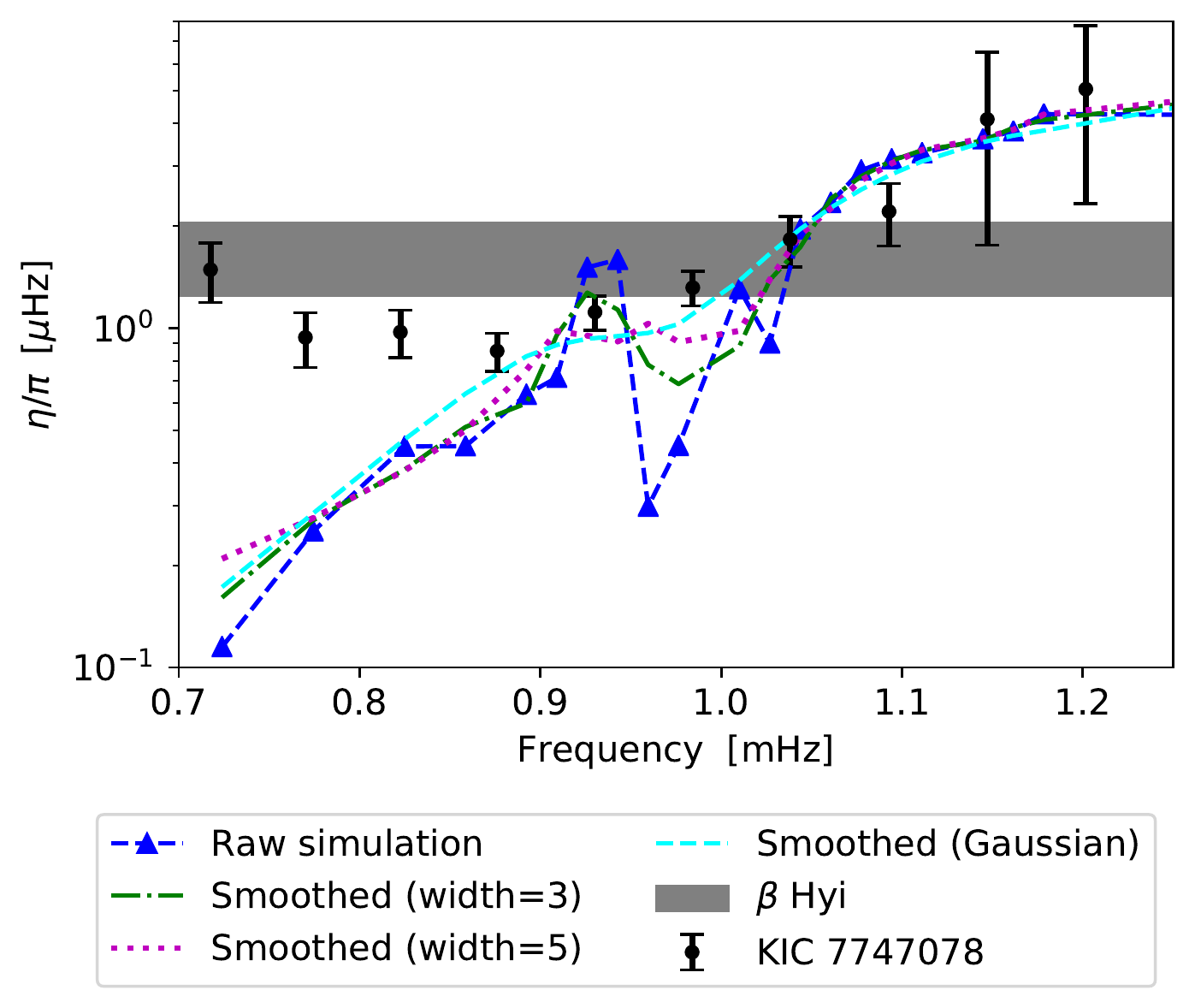}
\end{overpic}
\caption{Theoretical damping rates for $\beta$ Hydri. The cyan dashed line is smoothed from raw simulation data using a Gaussian kernel with an FWHM equal to 0.1 mHz. Mean line width for $\beta$ Hydri is from \citet{2007ApJ...663.1315B}, and KIC 7747078 is a subgiant whose basic parameters are close to $\beta$ Hydri (line width from \citealt{2020arXiv200506460L}).}
\label{fig:t59g40m00eta}
\end{figure}

\begin{figure}
\begin{overpic}[width=\columnwidth]{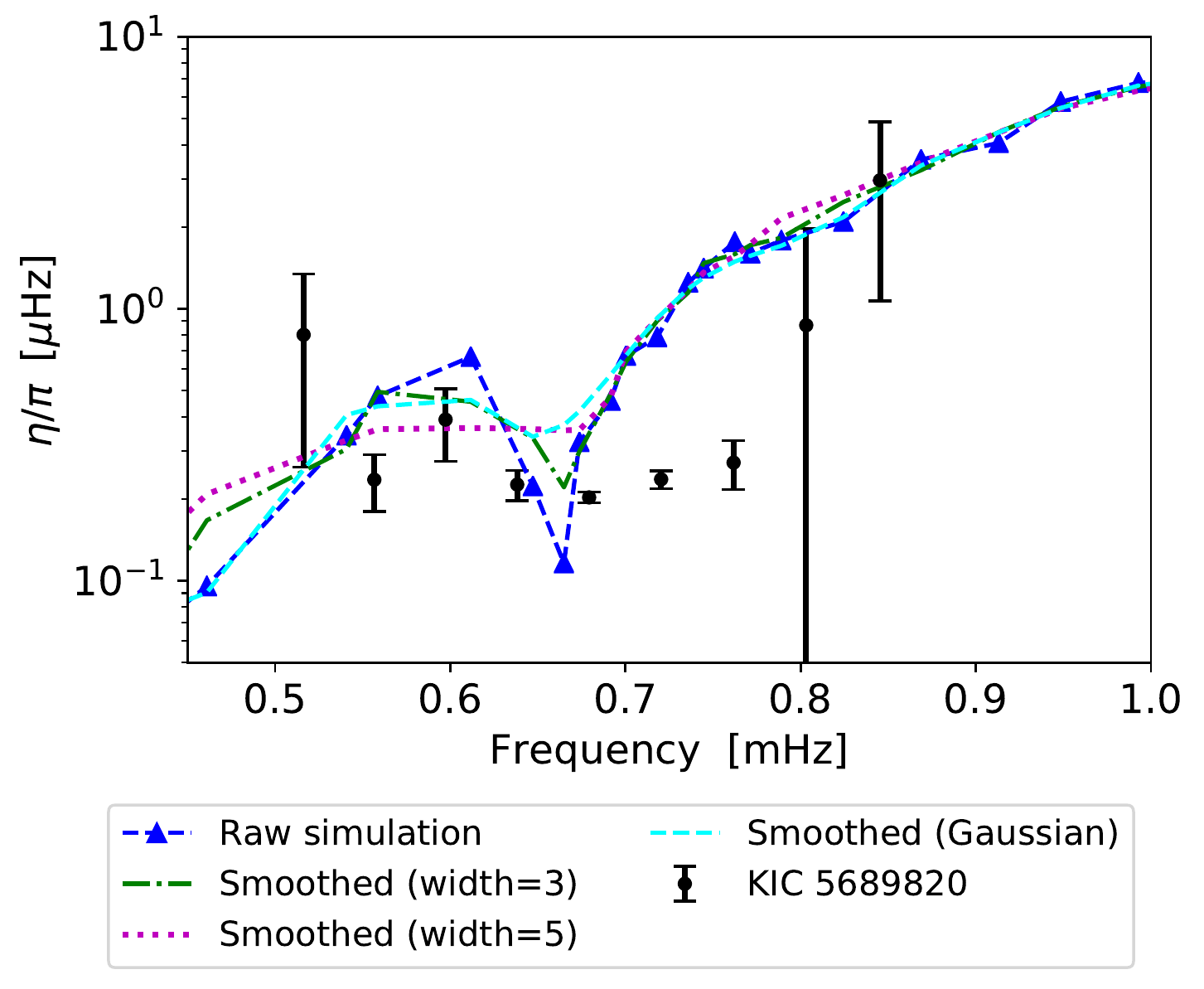}
\end{overpic}
\caption{Predicted damping rates for $\delta$ Eri are compared with measured line width of KIC 5689820 \citep{2020arXiv200506460L}, a star that has similar stellar parameters as $\delta$ Eri. The cyan dashed line is smoothed from raw simulation data using a Gaussian kernel with an FWHM equal to 0.07 mHz.}
\label{fig:t50g38m00eta}
\end{figure}


\begin{figure}
\begin{overpic}[width=\columnwidth]{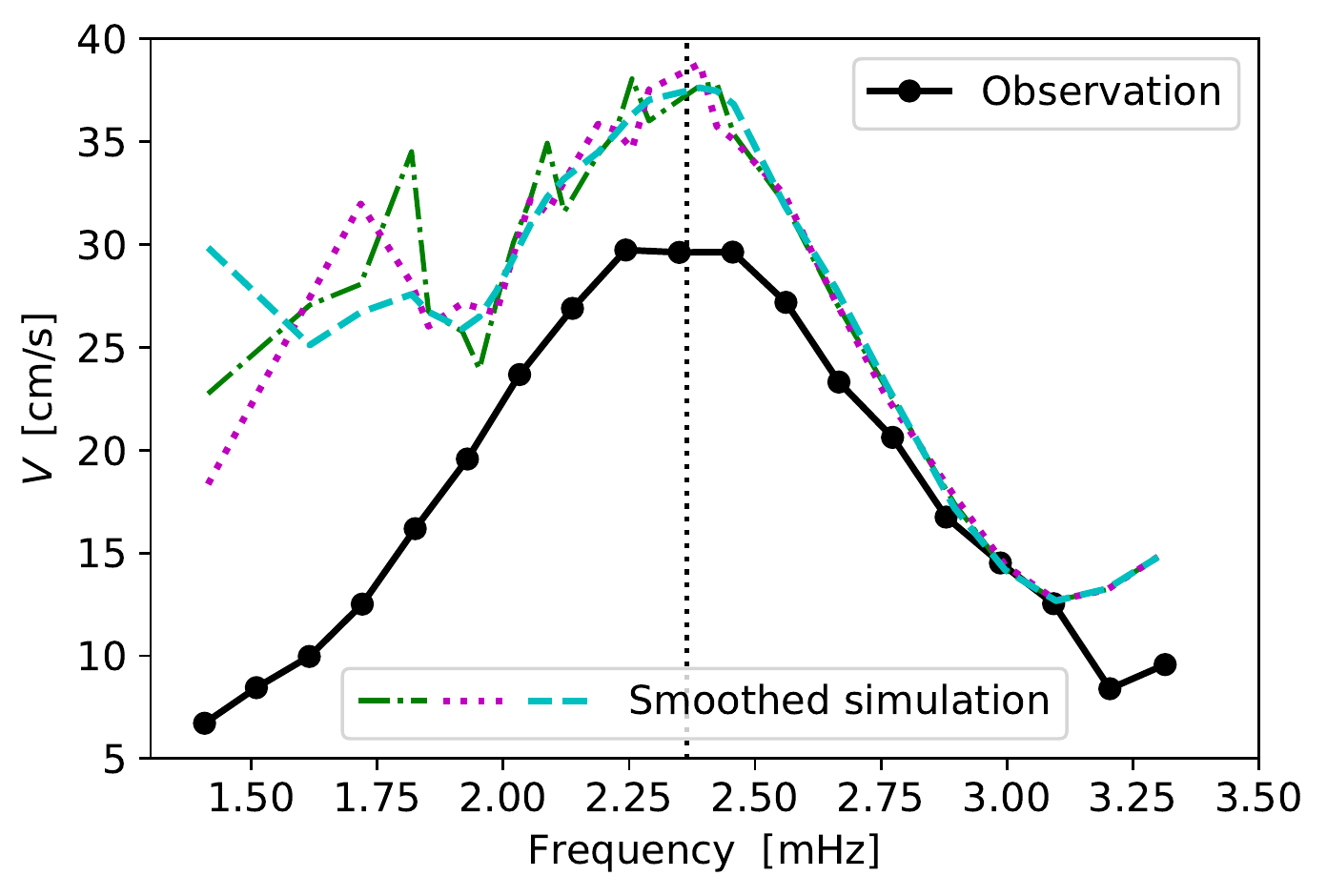}
\end{overpic}
\caption{Predicted photosphere velocity amplitude for KIC 6225718, as evaluated using Eq.~\eqref{eq:Vamp}. The green dashed-dotted, magenta dotted, and cyan dashed lines represent theoretical results from different smoothing options for damping rates (see Fig.~\ref{fig:t62g43m00eta}). Black dots represent the estimated velocity amplitude  for the same star converted from the observed flux variations \citep{2017ApJ...835..172L} using the empirical relation in \citet{1995A&A...293...87K}. The black vertical dotted line marks the observed $\nu_{\max}$.}
\label{fig:t62g43m00V}
\end{figure}

\begin{figure}
\begin{overpic}[width=\columnwidth]{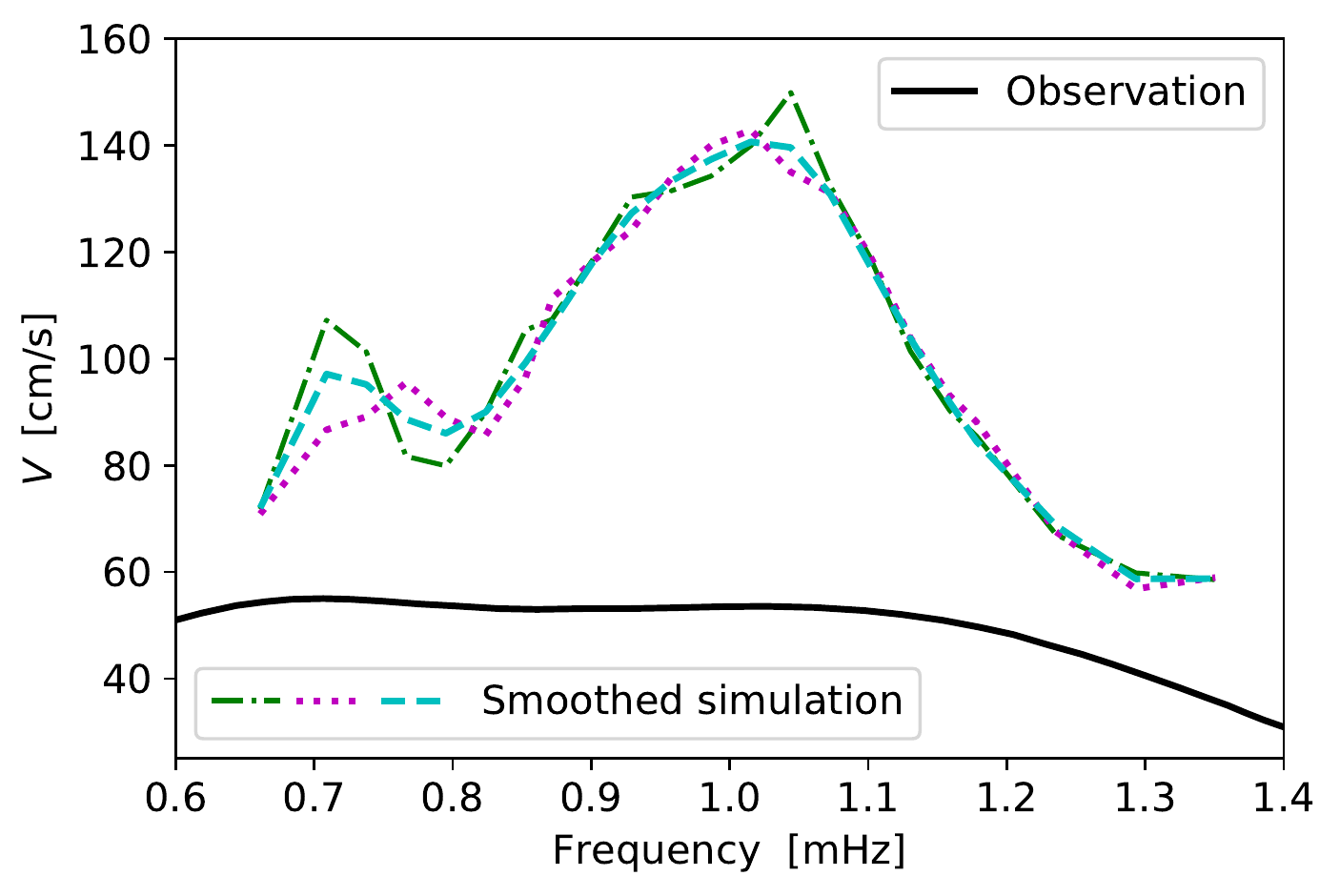}
\end{overpic}
\caption{Similar to Fig.~\ref{fig:t62g43m00V}, but photosphere velocity amplitude for Procyon. The smoothed mean radial velocity, measured by \citet[black solid line]{2008ApJ...687.1180A}, has been divided by the projection factor 0.712 in order to convert to kinematic velocity amplitude.}
\label{fig:t66g40m00V}
\end{figure}

\begin{figure}
\begin{overpic}[width=\columnwidth]{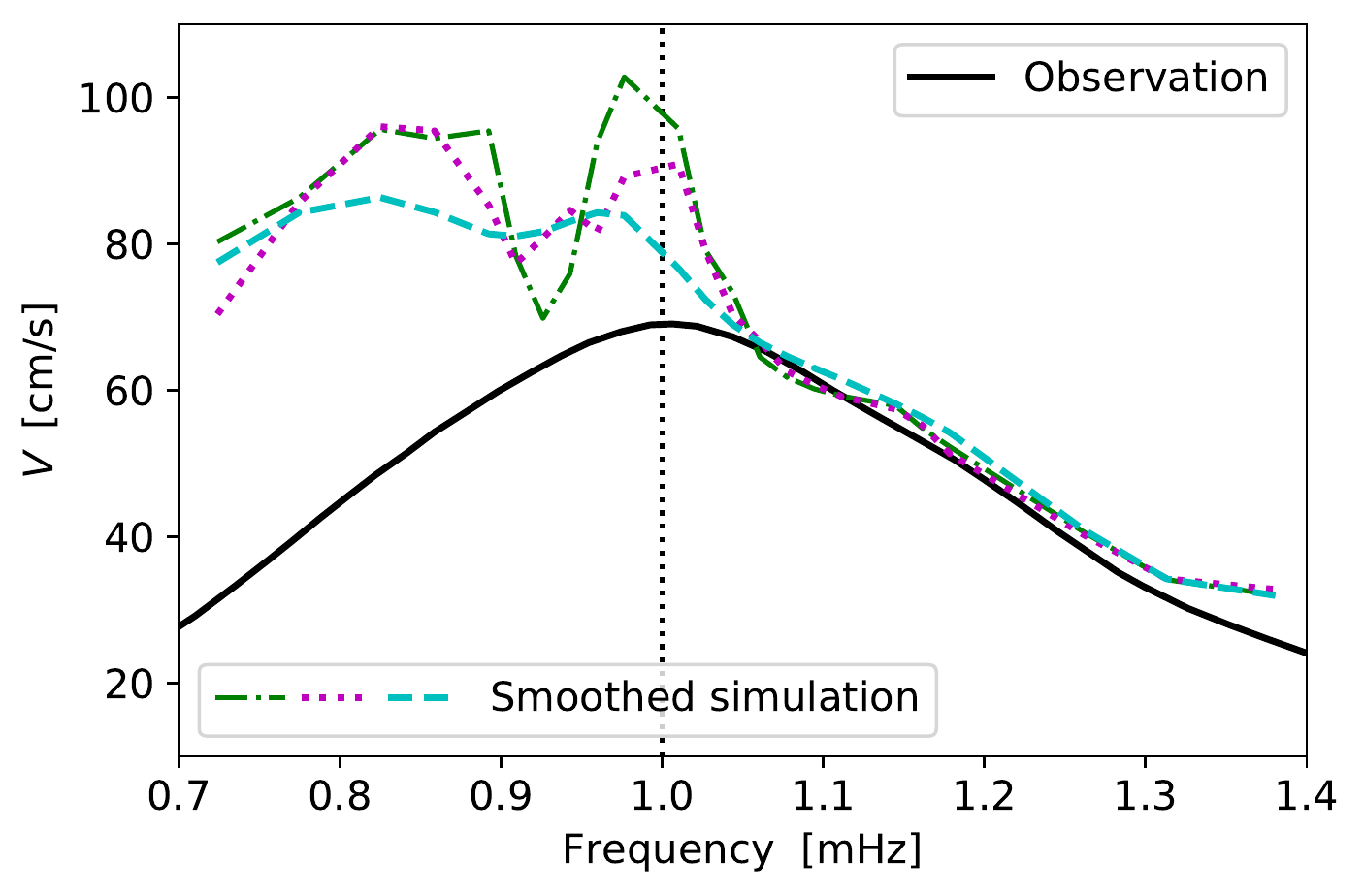}
\end{overpic}
\caption{Predicted photosphere velocity amplitude for $\beta$ Hydri is compared with measured mean radial velocity by \citet{2007ApJ...663.1315B}.}
\label{fig:t59g40m00V}
\end{figure}

\begin{figure}
\begin{overpic}[width=\columnwidth]{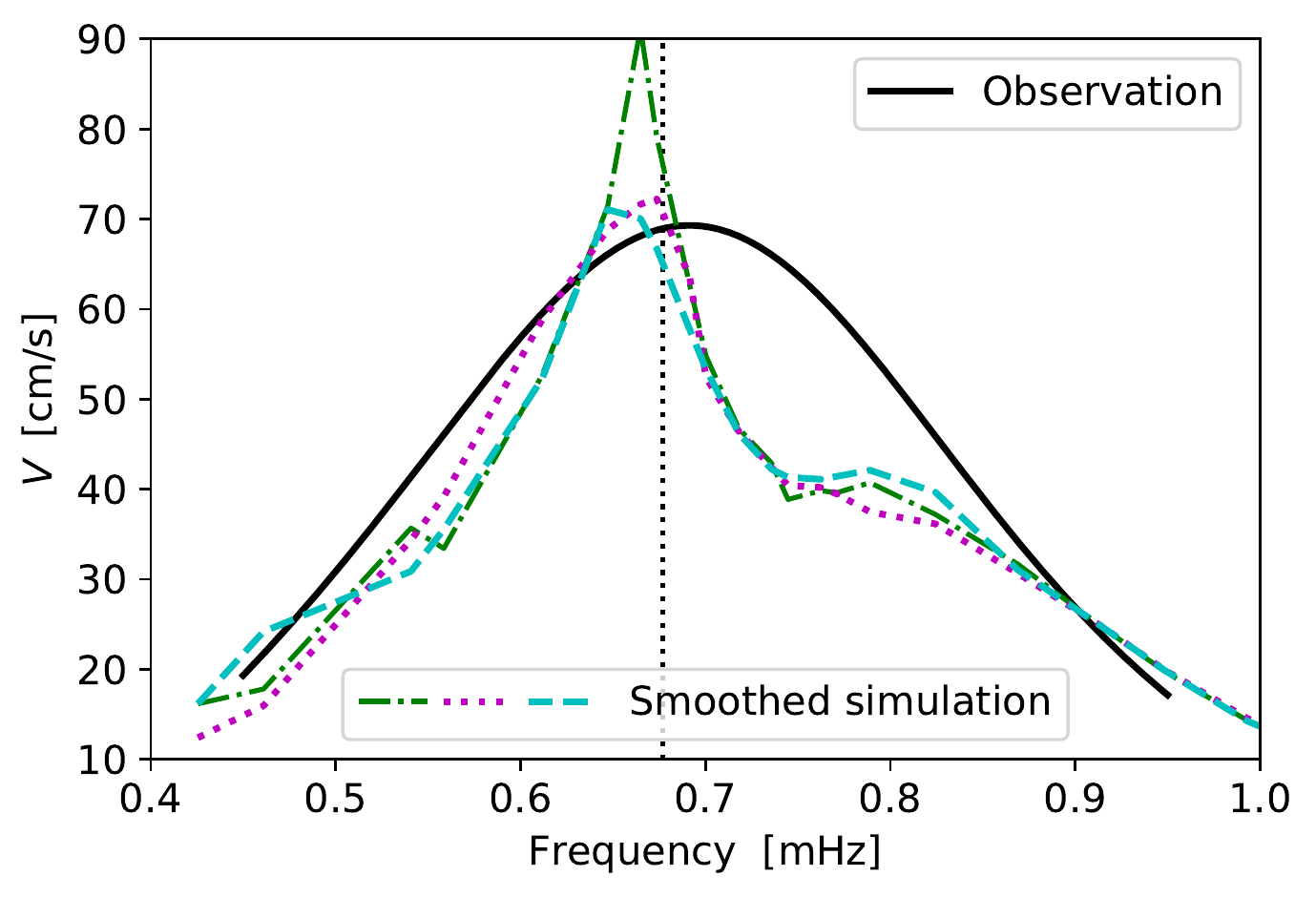}
\end{overpic}
\caption{Similar to Fig.~\ref{fig:t59g40m00V} but for $\delta$ Eri. Mean radial velocity (black solid line) is measured by SONG and data provided by E.~Bellinger and T.~Arentoft.}
\label{fig:t50g38m00V}
\end{figure}

Solar-like oscillations are p-modes driven by near-surface convection: fluctuations of thermodynamic quantities and turbulence caused by convection stochastically excite normal modes of the star to finite amplitude. Meanwhile, solar-like oscillations are dissipated by the same mechanisms that excite them \citep{2015LRSP...12....8H}. 
The final mean oscillation amplitude results from the balance between energy injection (excitation) rate and energy dissipation (damping) rate. In this section, we quantify both the excitation and damping rate of radial oscillations for our target stars from the theoretical angle. Mode excitation and damping rates are computed from 3D atmosphere and 1D patched models, which then allows an estimation of mode amplitude and $\nu_{\max}$.
  
Excitation rates are calculated based on Eq.~16 of \citet{2019ApJ...880...13Z}\footnote{$\mathfrak{Re}\lbrace f \rbrace$ ($\mathfrak{Im}\lbrace f \rbrace$) means the real (imaginary) part of complex function $f$.},
\begin{equation} \label{eq:er_num}
\begin{aligned}
\mathcal{P}_{\rm exc}(\omega) =&
\frac{\omega^2 A_{\rm box}}{8 \mathcal{T}_{\rm tot} E_0} \left[ 
\left( \int_{r_{\rm 3D \: bot}}^{r_{\rm surf}} \frac{\partial \xi_r}{\partial r} 
\mathfrak{Re}\left\lbrace \mathcal{F}[\delta \bar{P}_{\rm nad}] \right\rbrace \: dr \right)^2  \right.
\\
&+ \left. \left(\int_{r_{\rm 3D \: bot}}^{r_{\rm surf}} \frac{\partial \xi_r}{\partial r} 
\mathfrak{Im}\left\lbrace \mathcal{F}[\delta \bar{P}_{\rm nad}] \right\rbrace \: dr \right)^2 \right],
\end{aligned}
\end{equation}
where $\mathcal{F}$ represents the Fourier transform from time to frequency domain. The terms $\omega$, $A_{\rm box}$ and $\mathcal{T}_{\rm tot}$ stand for angular frequency, horizontal area of 3D simulation, and total time duration of 3D simulation, respectively. The term $E_0$ represents mode kinetic energy per unit surface area (\citealt{2001ApJ...546..576N} Eq.~63; \citealt{2019ApJ...880...13Z} Eq.~17): 
\begin{equation}
E_0 = \frac{\omega^2}{2} 
\int_0^{r_{\rm surf}} \rho \xi_r^2(r) \left(\frac{r}{R_{\rm phot}}\right)^2 \: dr,
\end{equation}
which is constant at given frequency. Here $\xi_r$ is the radial amplitude function (also called mode eigenfunction). It is calculated from 1D patched model using \textsc{adipls}. Its gradient, $\partial_r \xi_r$, represents the local compression of the fluid due to oscillations. $\delta \bar{P}_{\rm nad}$ is the horizontally averaged non-adiabatic pressure fluctuation, which includes all non-adiabatic effects such as entropy fluctuation and turbulence (Reynolds stress) caused by convection \citep{2001ApJ...546..576N}.
The time-dependent non-adiabatic pressure fluctuation is computed from the 3D simulation, then transferred to frequency space for the evaluation of excitation rate $\mathcal{P}_{\rm exc}$. Eq.~\eqref{eq:er_num} is integrated from the bottom of simulation domain $r_{\rm 3D \: bot}$ to the uppermost point of the patched model $r_{\rm surf}$, as $\delta \bar{P}_{\rm nad}$ is accessible in practice only through the 3D simulation. 
Eq.~\eqref{eq:er_num} implies that mode excitation results from the coupling between oscillations and convection. Excitation rates as a function of frequency are shown in Fig.~\ref{fig:er} for the four target stars. We refer the reader to \cite{2001ApJ...546..576N}, \cite{2001ApJ...546..585S} and \citet{2019ApJ...880...13Z} for detailed derivation of Eq.~\eqref{eq:er_num} and explanations about how components of this equation are computed numerically.

  The dissipation of oscillation energy is quantified by the damping rate $\eta$, which describes how fast an oscillation mode looses its kinetic energy by a factor of $e$ if there is no external energy supply. Damping processes broaden the power spectrum of the mode, shaping it to a Lorentzian profile centred at the eigenfrequency of the mode. The width of the Lorentzian envelope (line width $\Gamma$), an observable in asteroseismology, is connected to the damping rate by $\Gamma = \eta / \pi$ if the observational time series is much longer than the mode lifetime \citep{2005MNRAS.360..859C}. Throughout the paper, we confine our discussions to linear damping rates, which are derived assuming non-adiabatic effects can be treated as first-order perturbation to adiabatic oscillations. Nonlinear interactions are not likely to contribute significantly to the damping of radial modes for stars investigated in this work, according to the results from \citet{1994ApJ...427..483K}. The expression of (linear) damping rate for radial oscillations is:
\begin{equation} \label{eq:etamain}
\eta = \frac{\omega \int_{y_{\rm bot}}^{y_{\rm top}}
 \mathfrak{Im}\left\{ (\delta\bar{\rho}^{*} / \bar{\rho}_0) \delta \bar{P}_{\rm nad} \right\} dy }{4 m_{\rm mode} \vert \mathcal{V}(R_{\rm phot})\vert^2}
\end{equation}
(see Appendix \ref{sec:etath} for derivation). Here, the asterisk represents the complex conjugate, and $m_{\rm mode}$ and $\mathcal{V}$ are mode mass per unit surface area and vertical velocity amplitude, respectively. The denominator is proportional to the kinetic energy of the mode. The integral in the numerator is the work integral, which is proportional to the energy loss rate of the mode. In practice, this is evaluated from the bottom ($y_{\rm bot}$) to the top ($y_{\rm top}$) of the simulation domain along the vertical ($y$) direction, because outside the 3D simulation domain $\delta P_{\rm nad}$ is unobtainable. The value of the work integral is determined by the magnitude of density fluctuation $\delta\rho$ and the non-adiabatic pressure fluctuation, as well as the phase difference between them; further detail about how components of Eq.~\eqref{eq:etamain} are computed can be found in Appendix \ref{sec:etanum}. Theoretical damping rates, both raw simulation data and smoothed results, are divided by $\pi$ in order to compare directly with measured radial mode line widths, as depicted in Figs.~\ref{fig:t62g43m00eta}-\ref{fig:t50g38m00eta} for the four target stars. In all cases, the Gaussian kernel used to smooth the raw simulation data has an FWHM $\approx 1.75 \Delta\nu$. 
We note that only mean line width (converted from the mean mode lifetime $t_{\rm mode} = \eta^{-1}$ derived in \citealt{2010ApJ...713..935B}) is available for Procyon. In order to facilitate more detailed comparison, we also include frequency-dependent line width data of KIC 12317678 from the \textit{Kepler} LEGACY sample \citep{2017ApJ...835..172L} in Fig.~\ref{fig:t66g40m00eta}. The basic parameters of KIC 12317678 are $T_{\rm eff} = 6580 \pm 77$ K, $\log g = 4.048_{-0.008}^{+0.009}$ dex, $\rm [Fe/H] = -0.28 \pm 0.1$ dex \citep{2017ApJ...835..172L}, suggesting that this star is similar to Procyon so it is likely to be comparable to our simulation results. The situation for $\beta$ Hydri is similar: individual mode line widths are currently not available. Therefore, we also show the observed frequency-dependent line widths of a \textit{Kepler} subgiant with similar fundamental stellar parameters (KIC 7747078, $T_{\rm eff} = 5903 \pm 74$ K, $\log g = 3.90 \pm 0.01$ dex, $\rm [Fe/H] = -0.22 \pm 0.15$ dex; stellar parameters, mode frequencies and line width data provided by Yaguang Li, private communication) for comparison. For $\delta$ Eri, currently no line width information is available and therefore we compare our theoretical results with radial mode line widths of the similar star KIC 5689820.

The balance between stochastic excitation and mode damping dictates the final mean amplitude of the mode. With the excitation and damping rate both quantified, the mean kinematic velocity amplitude at the photosphere due to one oscillation mode can be evaluated via
\begin{equation} \label{eq:Vamp}
V = \sqrt{ \frac{2\mathcal{P}_{\rm exc}}{M_{\rm mode} \eta} }
\end{equation}
(\citealt{2019ApJ...880...13Z} Eq.~25), where $M_{\rm mode}$ is mode mass defined in \cite{2010aste.book.....A} Eq.~3.140 (not to be confused with mode mass per unit area $m_{\rm mode}$). 
The excitation and damping rates used in Eq.~\eqref{eq:Vamp} come from smoothed, rather than raw, simulation data in order to mitigate the effects of random fluctuations found in the latter and make the theoretical $V$ more comparable with observations (note that the published observed radial velocity power spectra have already been smoothed to ensure the extracted oscillation amplitudes are independent of the stochastic effects of the mode excitation and damping).
The thus computed kinematic velocity amplitudes for KIC 6225718, Procyon, $\beta$ Hydri and $\delta$ Eri are shown in Figs.~\ref{fig:t62g43m00V}-\ref{fig:t50g38m00V}, respectively. 

  However, what is obtained from the spectroscopic measurements of stellar oscillations is not $V$ directly, but the mean radial velocity $\mathfrak{v}$, whose physical source is kinematic velocity but which is also impacted by limb darkening and other geometric effects. The relation between radial and kinematic velocity is quantified by the projection factor, which depends on the mode quantum number and the wavelength at which the spectral line is measured. \citet{1996MNRAS.280.1155B} and \citet{2008ApJ...682.1370K} have calculated the projection factor for radial oscillations, measured at 550 nm wavelength, to be 0.712. We adopt this value for all stars included in this work, i.e.~$\mathfrak{v} = 0.712 V$, to make comparison between simulation and observation possible.
Meanwhile, we note that for KIC 6225718, stellar oscillations are identified by measuring brightness changes using photometry. In this scenario, the asteroseismic observable is the flux variation representing the change in surface temperature induced by oscillations. In view of this, we convert the observed flux variation to radial velocity using the empirical relationship proposed by \citet[their Eq.~5]{1995A&A...293...87K}, then divide the estimated radial velocity by 0.712 to compare directly with our simulation results.

\section{Discussion} \label{sec:discuss}

  Our results presented in Sect.~\ref{sec:mode} not only provide insights into the physical processes responsible for the driving and damping of radial oscillations for individual stars, but also allow comparison among different types of stars.
  In Sect.~\ref{sec:compobs}, we compare the theoretical damping rates and velocity amplitudes with observation and estimate theoretical $\nu_{\max}$ for our sample stars. We then discuss the connection between mode excitation/damping and global properties of stars (Sect.~\ref{sec:compglo}), and explore how radial oscillations are damped in the near-surface region of the star based on simulation results (Sect.~\ref{sec:damp}).

\subsection{Does 3D surface convection simulations agree with observation?} \label{sec:compobs}

  The theoretical damping rates for KIC 6225718 agree with observational data in general, as shown in Fig.~\ref{fig:t62g43m00eta}. The observed damping rates demonstrate a dip around 2.2 mHz, which is also predicted in the simulation results, albeit at slightly lower frequencies. However, below 1.9 mHz, we systematically underestimate the damping rates, with the discrepancy becoming larger toward lower frequencies. This misalignment is associated with the limited vertical size of the 3D simulation. As also pointed out in \citet{2019ApJ...880...13Z}, because the work integral is truncated at the bottom of the simulation box, contributions from deeper layers are omitted. This has greater influence on the low-frequency radial modes, as they have more substantial oscillation amplitudes in the deep stellar interior than high-frequency ones. We demonstrate this effect explicitly in Sect.~\ref{sec:etay}. 
  According to Eq.~\eqref{eq:Vamp}, the magnitude of the damping rates has a direct impact on the velocity amplitudes. For KIC 6225718, good agreement between theoretical and observationally inferred velocity amplitudes is achieved above 2 mHz, as seen from Fig.~\ref{fig:t62g43m00V}. However, below 2 mHz, we over-predict the velocity amplitude, which is a consequence of underestimating the damping rates in this frequency range. The errors below 2 mHz prevent us from obtaining a clear bell-shape $V - \nu$ curve that resembles observation. Nevertheless, the synthesized velocity amplitudes clearly show a local peak located between 2.25 and 2.45 mHz, which enables an estimate for theoretical $\nu_{\max}$. We are aware that the exact value of theoretical $\nu_{\max}$ is somewhat ambiguous because it depends on how the raw simulation data is smoothed. To this end, only an estimated theoretical $\nu_{\max}$ value is provided. In the case of KIC 6225718, $\nu_{\max}$ obtained from 3D simulations is in the vicinity of 2.3 mHz which is consistent with the measured value $\nu_{\rm max,obs} = 2.364$ mHz.

  For Procyon, theoretical damping rates are compared with observation in Fig.~\ref{fig:t66g40m00eta}. Between 0.75 mHz and 1.4 mHz, our results fall nicely in the uncertainty range of observationally inferred mean damping rate, indicating a general consistency between modelling and observation. Meanwhile, damping rates predicted from the 3D simulations demonstrate two noticeable features that differ from the other three stars investigated in this work. First, above 0.8 mHz, $\eta$ is nearly constant with frequency. Second, the depression of $\eta$ in a certain frequency range, which is a common characteristic of solar-like oscillations, is not recognizable for Procyon. 
  These features are likely to be physically real rather than caused by numerical errors because a similar trend is also seen for the measured damping rates of KIC 12317678, whose basic stellar parameters are close to those of Procyon. The underlying reason for the absence of the depression in $\eta$ will be investigated in Sect.~\ref{sec:damp}. The predicted velocity amplitudes, however, are systematically higher than the observed values, especially between 0.9 mHz and 1.1 mHz, where $V$ is overestimated by a factor of 2 (Fig.~\ref{fig:t66g40m00V}). Since the damping rates are consistent with observation overall, this disagreement stems from the excitation rate which is likely to be over-predicted. The reason for this will be investigated further in future work.
  
  In the case of $\beta$ Hydri, encouraging agreement between the modelled and measured mean damping rate are attained, as shown in Fig.~\ref{fig:t59g40m00eta}. The simulations predict a dip in $\eta$ located between 0.95 mHz and 1 mHz, which is reasonable as the depression of the damping rate commonly appears near $\nu_{\max}$ of the star. Comparing with the frequency-dependent damping rates of the similar subgiant KIC 7747078, we find that our predictions resemble observations at high frequencies ($\nu \gtrsim 1$ mHz) but are underestimated in the low-frequency regime ($\nu \lesssim 0.9$ mHz).
  The underlying reason is the same as in the case of KIC 6225718: limited vertical coverage of 3D simulation truncates the work integral. The errors on $\eta$ at low frequencies then propagate into the theoretical velocity amplitude, resulting in higher values than what are measured from observation (Fig.~\ref{fig:t59g40m00V}). Nevertheless, it is still possible to make an estimation of theoretical $\nu_{\max}$ from the local peak of $V$ near 1 mHz. We conclude that $\nu_{\max}$ of $\beta$ Hydri predicted from numerical simulations resides in the neighbourhood of 0.98 mHz, which conform with observation.

  For $\delta$ Eri, the calculated damping rates agree reasonably well with observations (note that here theoretical $\eta$ are compared with observations from a similar star, rather than $\delta$ Eri itself). The predicted dip is located at 0.65 mHz, which is consistent with the observed dip. The main discrepancy between simulation and observation takes place between 0.7 mHz and 0.8 mHz, where theoretical results are larger than measured values for reasons that are not entirely clear. In this case, the discrepancy is likely not due to the time duration or limited vertical scale of the artificial driving simulations, because we have verified that (1) doubling the simulation timespan does not affect the damping rate result noticeably, and (2) the contribution from the deep layers of the simulation to the work integral is small (especially for $\nu \gtrsim 0.7$ mHz artificial driving simulations), implying that the vertical size of simulation is sufficient for modelling mode damping in this frequency range.  
Turning to velocity amplitude, our theoretical results generally agree with observations both in magnitude and in the shape of the $V - \nu$ curve. The predicted $\nu_{\max}$, which is estimated to be around 0.65 mHz, also matches the observed $\nu_{\max}$ (0.677 mHz) for $\delta$ Eri.
 
  When comparing simulation results with observations, one should keep in mind that magnetic fields, which are ubiquitous in stars but not included in our simulations, do interact with acoustic oscillations. Helioseismic analysis of low degree p-modes over the solar cycle \citep{2000MNRAS.313...32C} has demonstrated that mode excitation is not sensitive to magnetic fields. Damping rates (line widths) however, increase with increasing magnetic activity\footnote{A crude explanation is that granules become smaller as magnetic field strength increases (see \citealt{2009LRSP....6....2N} and references therein). The decreased granule size gives rise to larger damping rates near $\nu_{\max}$, according to the calculation by \citet{2001MNRAS.327..483H}.}. The net effect is that with increasing solar activity, the amplitude of solar p-modes decreases. The suppression of oscillation amplitude due to magnetic activity has also been confirmed in other solar-like oscillating stars \citep{2019A&A...628A.106B}. With this in mind, and noting that the signature of stellar activity was found in Procyon \citep{2011ApJ...731...94H}, it is worth discussing the potential influence of magnetic fields on our results. For the Sun, \citet{2000MNRAS.313...32C} measured a $\approx 25 \%$ increase in damping rates from solar activity minimum to maximum. Assuming the influence of stellar activity on damping rates and mode amplitudes in Procyon is qualitatively the same as the solar case, the presence of magnetic fields will results in slightly larger damping rates, and thus slightly smaller $V$ compared to predictions from our simulations.
  
  Although disagreements and uncertainties exist, the comparison between oscillation properties obtained from simulations against observations shows great promise. In all cases, encouraging agreement between the two is achieved. Without introducing any tunable parameter in our calculations, the computed damping rates and velocity amplitudes are of the same order of magnitude as the corresponding measured values, and in most cases, the $\eta - \nu$ and $V - \nu$ relations also resemble observation. Theoretical $\nu_{\max}$ values estimated from our simulations are consistent overall with the corresponding measurements. These indicate that our numerical approach for modelling the excitation, damping and amplitude of radial oscillations is applicable not only to the Sun (demonstrated in \citealt{2019ApJ...880...13Z}) but also to other solar-type oscillators.

\subsection{What is the relationship between excitation/damping rate and fundamental stellar parameters?} \label{sec:compglo}

\begin{figure}
\begin{overpic}[width=\columnwidth]{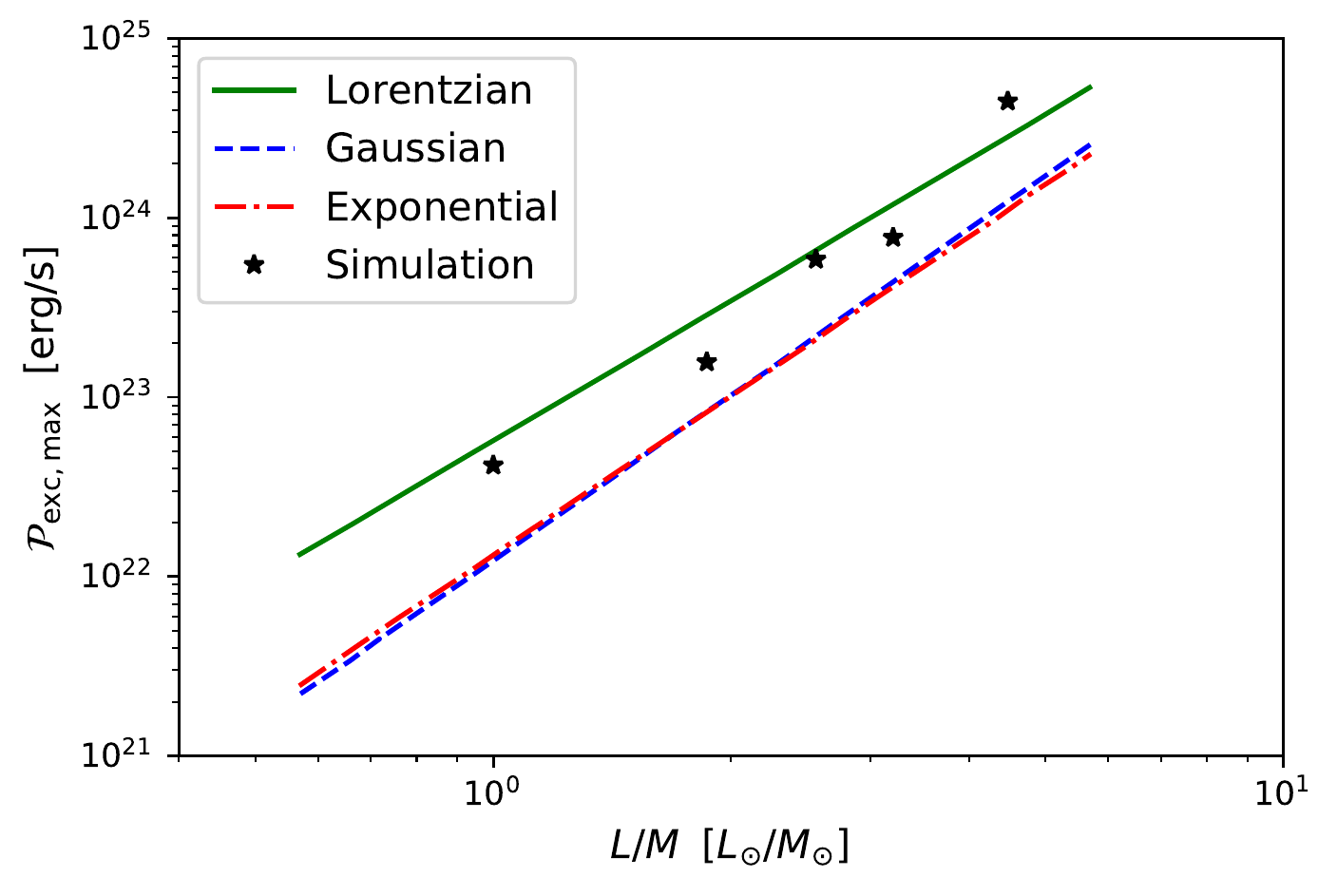}
\end{overpic}
\caption{The relationship between maximum excitation rate and luminosity-mass ratio of the star is shown. Solid, dashed and dashed-dotted lines are relations derived in \citet[their Fig.~6]{2007A&A...463..297S}, with label ``Lorentzian'', ``Gaussian'' and ``Exponential'' represent different analytical models for turbulence in their calculations (detailed in \citealt{2007A&A...463..297S} and references therein). Black star symbols are results from our simulations, where $\mathcal{P}_{\rm exc,max}$ are evaluated by taking the maximum value of the smoothed excitation rates (not the raw simulation data in order to avoid strong fluctuations) and $L/M$ are obtained from \textsc{mesa} models.
}
\label{fig:LMvsermax}
\end{figure}

We are now in a position to study the relationship between mode excitation/damping and fundamental parameters of these stars. We emphasize that our sample size is too small to establish a quantitative relation, but we can discuss the qualitative behaviour.
  
The trend in excitation rate is similar for all stars investigated: mode excitation is weak at low frequencies, then increases with frequency to a plateau that contains $\nu_{\max}$ before slightly declining at higher frequencies. 
The underlying reason is explained in, for example, \citet{2001ApJ...546..585S} and \citet{2019ApJ...880...13Z}. In brief, at low frequencies, relatively weak local compression caused by the low-frequency mode limits the excitation rate (i.e.~small $\partial_r \xi_r$ in Eq.~\eqref{eq:er_num}). At high frequencies, on the contrary, mode excitation is limited by convection because non-adiabatic pressure fluctuation decreases with increasing frequency (i.e.~small $\delta P_{\rm nad}$ in Eq.~\eqref{eq:er_num}).

We note that excitation rate is overall greater in hotter (higher $T_{\rm eff}$) stars, which is clearly observed by comparing the Sun and KIC 6225718, or $\beta$ Hydri and Procyon, as their surface gravities are similar. The correlation between $\mathcal{P}_{\rm exc}$ and $T_{\rm eff}$ can be understood by considering the heat transport near the photosphere: Stronger radiative cooling and larger convective flux near photosphere are required to transport more energy in hotter stars \citep{2004SoPh..220..229S}. Larger radiative and convective fluxes then give rise to greater entropy fluctuation and stronger velocity field, which directly results in more energy supply from convection to oscillations via the stochastic excitation mechanism.
The relation between $\mathcal{P}_{\rm exc}$ and global stellar parameters has been empirically quantified by \citet{2007A&A...463..297S}, who suggested that excitation rate should scale with the luminosity-mass ratio (essentially the same as $T_{\rm eff}^4 / g$), 
\begin{equation} \label{eq:er_scale}
\mathcal{P}_{\rm exc,max} \propto (L/M)^s, 
\end{equation}
where $\mathcal{P}_{\rm exc,max}$ is the maximum excitation rate of the star and $s$ is a slope to be fixed by fitting to numerical results. The linear relation between $\log \mathcal{P}_{\rm exc,max}$ and $\log(L/M)$ obtained in \citet{2007A&A...463..297S} from their semi-analytical calculations of excitation rates for difference stars, together with results from our 3D simulations, are demonstrated in Fig.~\ref{fig:LMvsermax}. Our numerical results obey this scaling law, indicating excitation rates evaluated in this work are consistent with \citet{2007A&A...463..297S}, although our method is radically different from theirs.

  Damping rates are believed to depend on global stellar parameters as well. The scaling relation for damping rates was first proposed by \citet{2009A&A...500L..21C}. Based on observational data and their pulsation calculations, they suggested damping rates near $\nu_{\max}$ should be proportional to the fourth power of the effective temperature. The positive correlation between $\eta$ near $\nu_{\max}$ and $T_{\rm eff}$ was subsequently confirmed by \citet{2012A&A...537A.134A} and \citet{2018A&A...616A..94V} for main-sequence, subgiant and red giants observed by \textit{Kepler}. Owing to the limited sample size and errors on theoretical damping rates near $\nu_{\max}$, we do not attempt to present a quantitative relation between $\eta$ and $T_{\rm eff}$. Nonetheless, by comparing two cooler stars ($\delta$ Eri and $\beta$ Hydri, Figs.~\ref{fig:t50g38m00eta} and \ref{fig:t59g40m00eta}) with the two hotter stars (KIC 6225718 and Procyon, Figs.~\ref{fig:t62g43m00eta} and \ref{fig:t66g40m00eta}), it is obvious that damping rates (near $\nu_{\max}$) predicted from simulations increase with effective temperature of the star, which qualitatively agrees with observations.

  The magnitude of the damping rate is determined in part by the work integral, and thus the density and non-adiabatic pressure fluctuations (Eq.~\eqref{eq:etamain}). As demonstrated in \citet[their Figs.~2 and 3]{2013A&A...560A...8M}, fluctuations in thermodynamic quantities are stronger in hotter stars because of the relatively larger convective velocity field. Their findings offers insights into the positive correlation between $\eta$ and $T_{\rm eff}$: in hotter stars, stronger fluctuations in density and pressure result in larger damping rates\footnote{We are aware that the explanation provided here might not cover the whole picture of the $\eta - T_{\rm eff}$ relation, because damping rate is not only affected by the strength of fluctuations in density and pressure but also by the phase difference between them, and the mode kinetic energy also plays a role (see Eq.~\eqref{eq:etamain}).}.
  It is worth noting that both excitation and damping rates are positively correlated with effective temperature, which is not surprising because solar-like oscillations are excited and damped by the same physical process: turbulent convection \citep{2015LRSP...12....8H,2019ApJ...880...13Z}. Therefore, their relationship with global stellar parameters should be similar.
  
  We now proceed to the relation between $\nu_{\max}$ and fundamental stellar parameters, one of the most important scaling relations in asteroseismology. As first suggested by \citet{1991ApJ...368..599B} and \citet{1995A&A...293...87K}, $\nu_{\max}$ of a solar-like oscillating star should scale with its surface gravity and effective temperature, $\nu_{\max} \propto g/\sqrt{T_{\rm eff}}$. By scaling from the solar values, the thus evaluated $\nu_{\max}$ are 2.26 mHz for KIC 6225718, 1.07 mHz for $\beta$ Hydri and 0.70 mHz for $\delta$ Eri, which broadly agree with our theoretical results. Given that theoretical $\nu_{\max}$ are estimated from \textit{ab initio} hydrodynamical simulations -- an approach completely independent of the $\nu_{\max}$ scaling relation -- we claim the overall validity of the $\nu_{\max}$ scaling relation is supported at solar-metallicity from a theoretical angle.

  In this work, we have successfully derived the relationship between excitation and damping rates and fundamental stellar parameters from a purely theoretical perspective: Excitation and damping rates near $\nu_{\max}$ of the star are both positively correlated with the effective temperature, consistent with previous theoretical investigations and empirical findings for solar-type oscillating stars. In addition, theoretical $\nu_{\max}$ estimated from simulations scales with $g/\sqrt{T_{\rm eff}}$, confirming qualitatively the $\nu_{\max}$ scaling relation at solar-metallicity. These findings suggest that our numerical approach for modelling the excitation and damping of radial modes is valid across a wide range of effective temperatures and surface gravities.
  With more detailed numerical simulations that cover additional $\{ T_{\rm eff}, \log g, \rm [Fe/H] \}$ combinations, especially including red giant branch stars, which are important in Galactic archaeology (e.g.~\citealt{2016MNRAS.455..987C}) but have not yet been studied with 3D models, it should be possible to quantify the relationship between $\mathcal{P}_{\rm exc}$, $\eta$ and fundamental stellar parameters and even to quantify the departure (if any) from the widely-used $\nu_{\max}$ scaling relation from 3D surface convection simulations.

\subsection{What is the underlying physics of mode damping?} \label{sec:damp}

\begin{figure*}
\subfigure[]{
\begin{overpic}[width=0.48\textwidth]{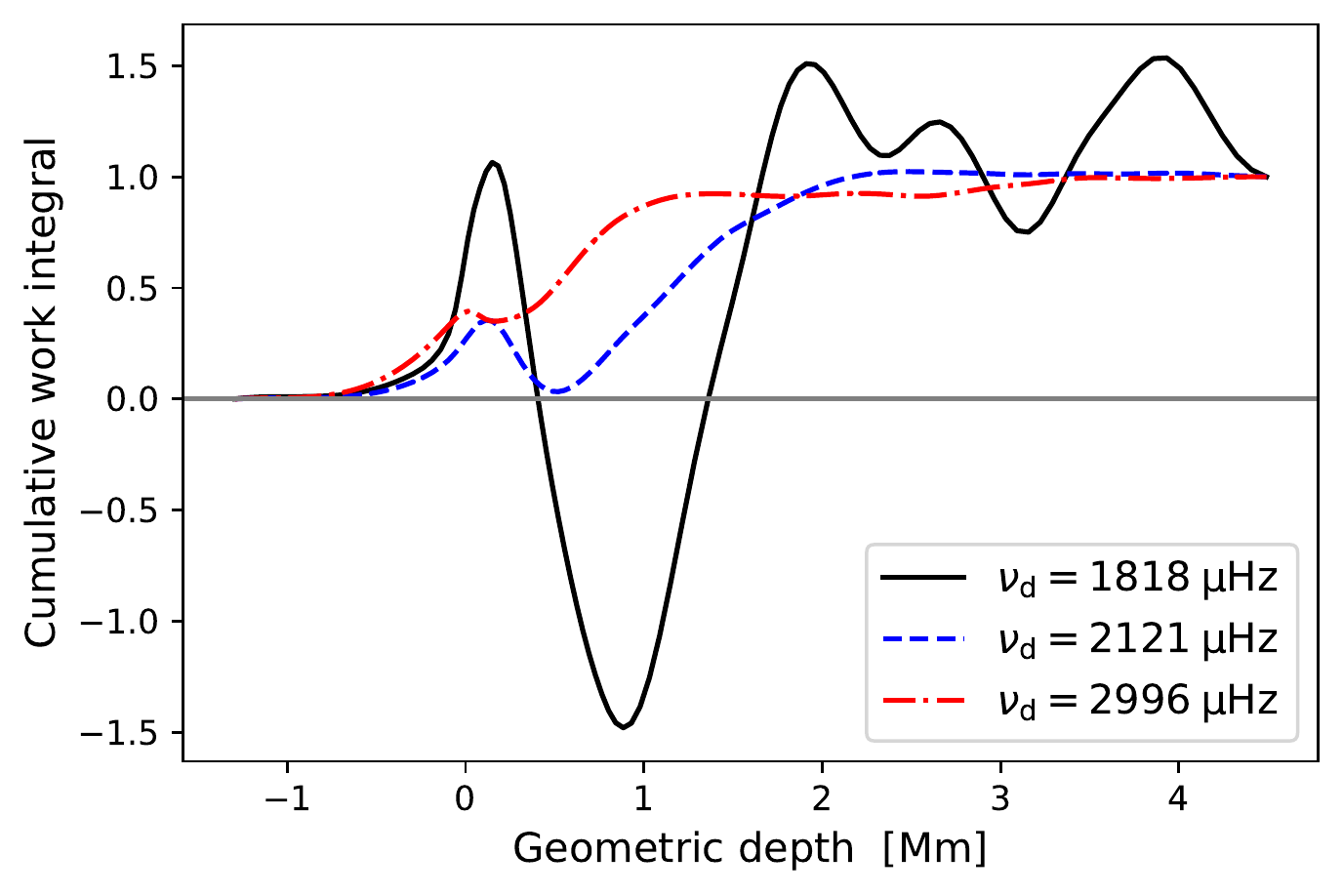}
\end{overpic}
\label{fig:work_mode} 
}
\subfigure[]{
\begin{overpic}[width=0.48\textwidth]{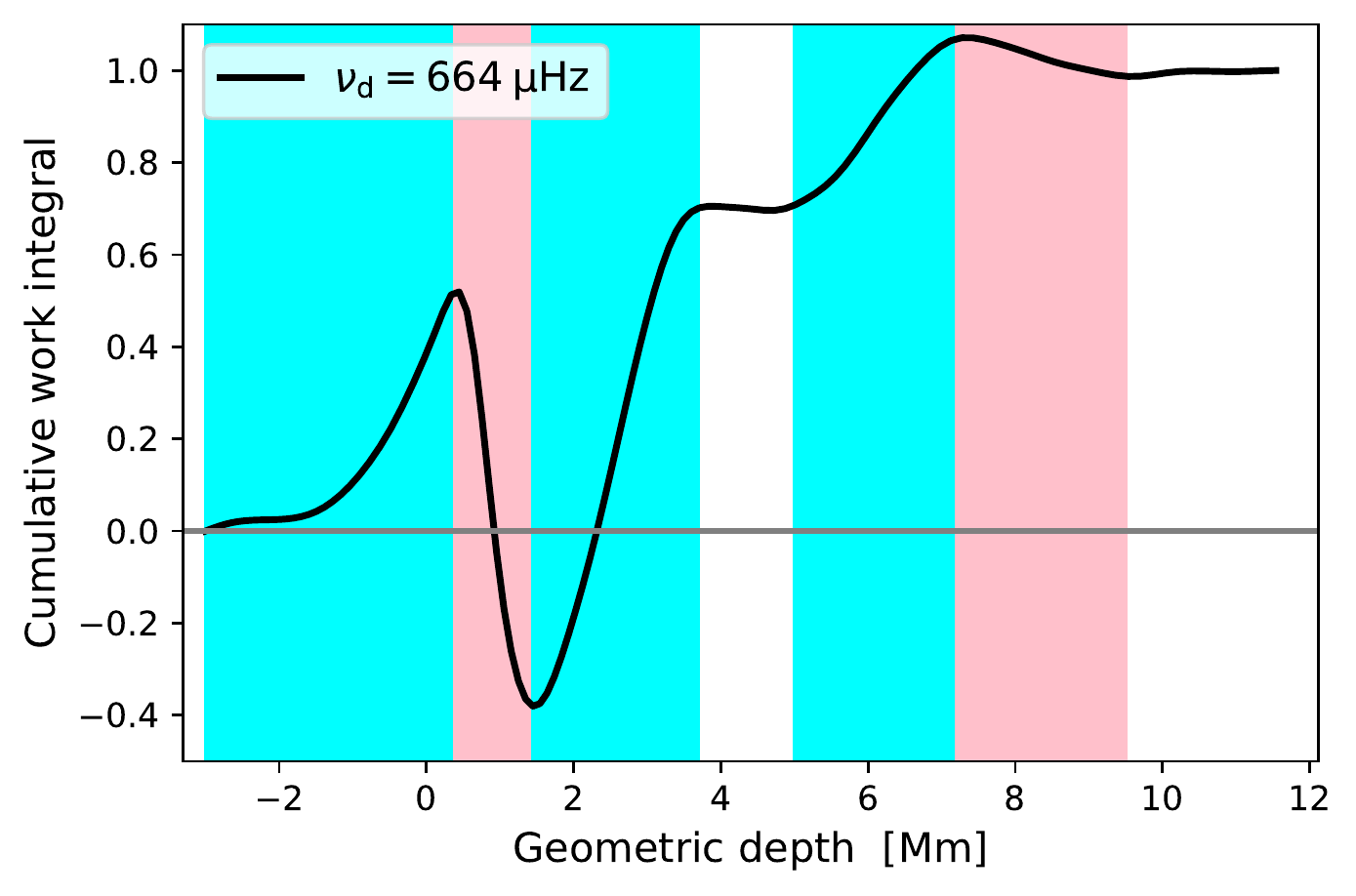}
\end{overpic}
\label{fig:work_demo} 
}
\caption{
\ref{fig:work_mode}: Normalized cumulative work integral distributions within the simulation domain for three example simulation modes, computed from artificial driving experiments for KIC 6225718 with driving frequency $\nu_{\rm d}$ equals to 1818 $\mu$Hz (low-frequency), 2121 $\mu$Hz (intermediate-frequency) and 2996 $\mu$Hz (high-frequency), respectively. 
Here, the work integral is integrated from the top (left side of the figure) to the bottom of the simulation domain and normalized by its total value, therefore at the top the cumulative work integral is 0 while at bottom it is always 1. Zero geometric depth corresponds approximately to the photosphere. 
\ref{fig:work_demo}: Similar to \ref{fig:work_mode}, but illustrating the damping and growth region of a simulation mode computed from an artificial driving experiment for $\delta$ Eri. Noticeable damping and growth areas are shaded in cyan and pink, respectively.
\label{fig:work_int} }
\end{figure*}

\begin{figure}
\begin{overpic}[width=\columnwidth]{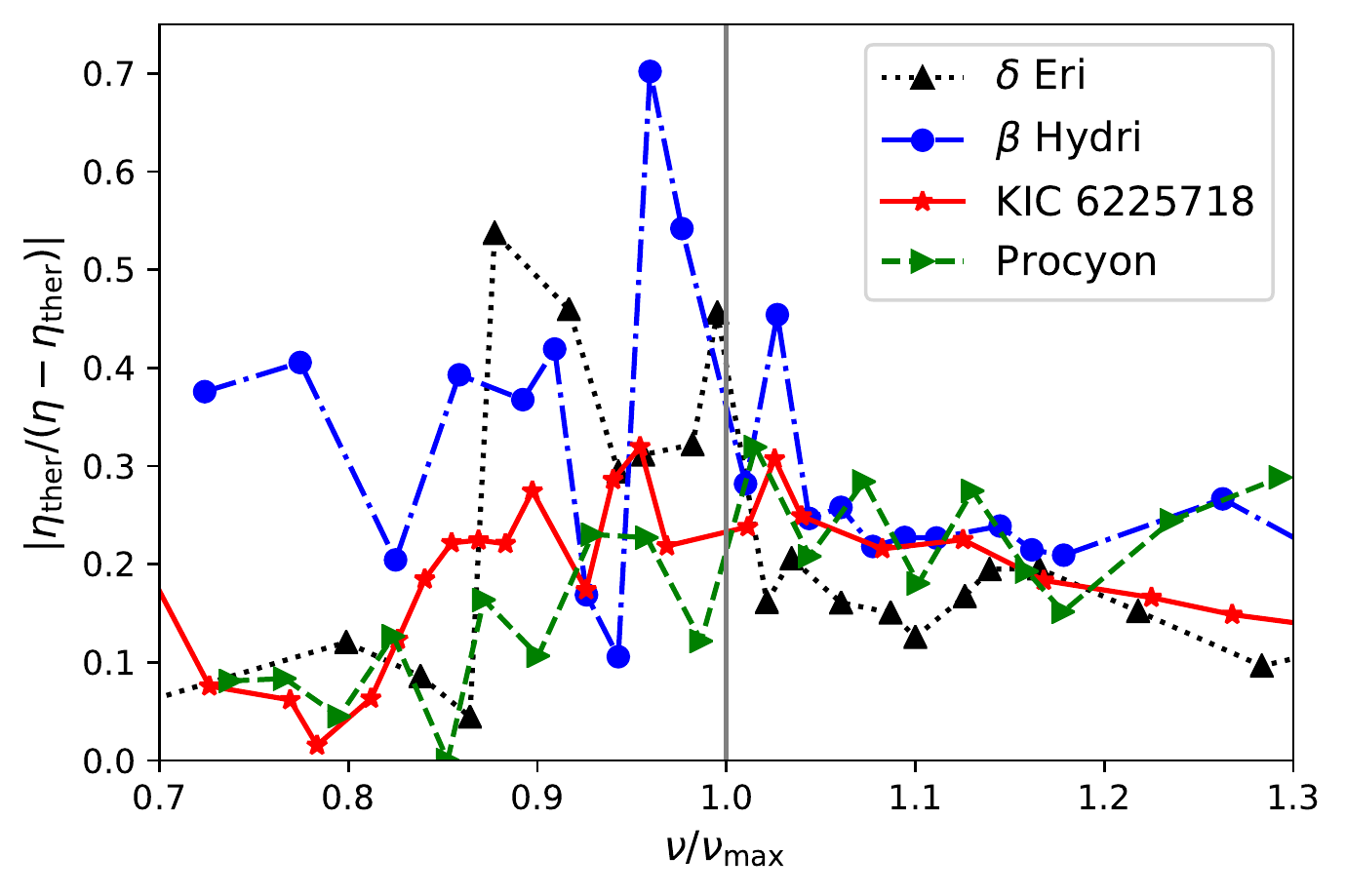}
\end{overpic}
\caption{The ratio between $\eta_{\rm ther}$ and $\eta - \eta_{\rm ther}$ at different frequencies for all four target stars. Frequencies are normalized by the measured $\nu_{\max}$ of the corresponding star (see table \ref{tb:seis_param}, note that for Procyon frequencies are normalized by 1 mHz because the value of $\nu_{\max}$ is uncertain for this star). All data presented here are computed from artificial driving simulations without any smoothing.
}
\label{fig:eta_ratio}
\end{figure}

  Apart from quantifying the value of the damping rates at different frequencies, it is also worthwhile to understand the physics behind mode damping, an important topic that is difficult to probe by observation. In this section, we will explore two relevant questions based on the simulation results: (1) Which part of the star contributes most to mode damping, and hence dictates the final value of $\eta$? (2) The dip in damping rate near $\nu_{\max}$ is a common feature in the $\eta - \nu$ curve, but why is it less pronounced in warm turn-off stars like Procyon?
  
  To answer question (1), we show in Fig.~\ref{fig:work_int} the cumulative work integral distribution in the entire simulation domain, which reflects contributions to damping from different locations in the atmosphere and upper convection zone. Increasing work integral (with geometric depth) means positive work is done by the mode at the corresponding location, suggesting the mode is damped there. Conversely, decreasing work integral means negative work, indicating that the mode is growing locally. Fig.~\ref{fig:work_demo} clearly demonstrates several damping and growth regions for an example simulation mode, and the relative strength between damping and growth determines the stability of this mode. Although modes (in the same star) with different frequencies are damped and driven in different regions, they have some features in common. As observed from Fig.~\ref{fig:work_mode}, all three modes shown are damped in the stellar atmosphere (negative geometric depth). Moving inward, there is a growth region just below the photosphere where temperature stratification is highly superadiabatic (see e.g.~Fig.~25 in \citealt{2013A&A...557A..26M}), implying tight connection between mode energy gain/loss and over-adiabaticity. Intermediate- and high-frequency modes are mostly damped in deeper layers, and their cumulative work integral becomes nearly flat when approaching the bottom of simulation domain. The latter indicates that the vertical size of the simulation box is sufficient for modelling damping processes for these two modes. The low-frequency mode, however, demonstrates broader regions of damping and growth, in agreement with \citet{1992MNRAS.255..603B} for low-frequency p-modes in the Sun. The fact that discernable regions of damping and growth are present down to the bottom of the simulation domain also suggests that extra contributions to the damping of the low-frequency modes from the deep interior are omitted because of the limited vertical size of simulation, as discussed in Sect.~\ref{sec:compobs}.

  Regarding question (2), Procyon is not an anomalous case showing an ``odd'' $\eta - \nu$ relation, but rather a typical representation of warm stars. Both observation \citep{2014A&A...566A..20A} and theoretical investigation \citep{2019MNRAS.487..595H} have confirmed that the dip in $\eta$ near $\nu_{\max}$ becomes less obvious with increasing $T_{\rm eff}$, and above $\sim 6300$ K it is hardly seen (demonstrated in Fig.~3 of \citealt{2014A&A...566A..20A} and Figs.~A1-A3 of \citealt{2019MNRAS.487..595H}). Therefore, a more appropriate question may be: why does the dip in $\eta$ disappear in warm ($T_{\rm eff} \gtrsim 6300$ K) stars?
To answer this, one should first understand the physical origin of the damping rate dip, which is explained in \citet{1992MNRAS.255..603B} for the solar case. In short, at frequencies where the damping rate dip occurs, destabilizing effects from the thermal pressure fluctuations largely cancel the stabilizing effects from convective turbulence, leaving a relatively small $\eta$ compared to lower or higher frequencies, where the cancellation is relatively less severe. 
Here, we show the relative importance of thermal processes (radiative and convective heat transport) and turbulence to mode damping by displaying $|\eta_{\rm ther} / (\eta - \eta_{\rm ther})|$ in Fig.~\ref{fig:eta_ratio} for four stars investigated in this work, where $\eta_{\rm ther}$ represents contributions to damping rates from thermal pressure fluctuations while $\eta - \eta_{\rm ther}$ mainly reflects damping due to turbulent pressure. Larger $|\eta_{\rm ther} / (\eta - \eta_{\rm ther})|$ thus signifies that thermal processes have a greater impact on the total damping rate. As illustrated in Fig.~\ref{fig:eta_ratio} for Procyon and KIC 6225718, $|\eta_{\rm ther} / (\eta - \eta_{\rm ther})|$ near $\nu_{\max}$ is typically less than 0.3, whereas in the other two cooler stars, it is $\sim 0.5$ near $\nu_{\max}$. Although our sample is not large enough to draw a definite conclusion, it is very likely that thermal processes are less influential in mode damping for warm stars. As thermal processes are responsible for destabilizing modes with frequency near $\nu_{\max}$, relatively large (relative to the contribution from turbulence) negative $\eta_{\rm ther}$ will hence depress $\eta$ near $\nu_{\max}$, which is the situation of $\delta$ Eri and $\beta$ Hydri. In warm stars, however, thermal processes are not significant enough to leave noticeable fingerprints on the $\eta - \nu$ curve, which is dominated by contributions from the convective motion.
  
  In addition, we clarify the connection between the mixing length parameter $\alpha_{\rm MLT}$ and the damping rate dip, which is discussed in \citet{1992MNRAS.255..603B} and \citet{2014A&A...566A..20A}. \citet{1992MNRAS.255..603B} has shown that the damping rate dip predicted for the Sun becomes less pronounced with increasing $\alpha_{\rm MLT}$. On the other hand, $\alpha_{\rm MLT}$ calibrated from 3D convection simulations decreases with increasing $T_{\rm eff}$ \citep{1999A&A...346..111L,2014MNRAS.445.4366T,2015A&A...573A..89M}. Given that the damping rate dip becomes less obvious with increasing $T_{\rm eff}$, both observationally and from our simulations, at first glance it seems these two conclusions contradict each other. However, both conclusions can in fact be valid. In the Sun, for example, increasing $\alpha_{\rm MLT}$ results in more efficient convective heat transfer. 
  That is, an equal amount of heat can be carried in a region with less over-adiabaticity, causing the superadiabatic temperature gradient to become smaller near the surface \citep{2018ApJ...856...10J}.
Consequently, the destabilizing effects from thermal processes decrease. The larger velocity field with increasing $\alpha_{\rm MLT}$ strengthens the contribution from convective turbulence to mode damping. These two factors together make thermal processes less significant, thus translating to a less pronounced damping rate dip when assuming larger $\alpha_{\rm MLT}$. Comparing stars with different $T_{\rm eff}$, we see that in warmer stars, the superadiabatic temperature gradient is larger near the photosphere, as predicted from 3D simulations \citep{2013A&A...557A..26M}. Therefore, the calibrated $\alpha_{\rm MLT}$ is smaller and thermal processes are stronger. Nevertheless, the velocity field and convective turbulence are also much stronger in warmer stars and the contribution from convective turbulence to mode damping increases with $T_{\rm eff}$ as well. As discussed above, it is the relative importance between thermal processes and convective turbulence that determines the shape of the $\eta - \nu$ curve. Since thermal processes are relatively less influential in hot stars, the dip in $\eta$ is less recognizable, which does not conflict with a comparatively small $\alpha_{\rm MLT}$.

\section{Conclusions}

  In this work, we quantified the excitation and damping rates of radial p-modes based on detailed modelling of both the atmospheres and interiors of four key benchmark stars exhibiting solar-like oscillations. We adopted the theoretical framework of \citet{2001ApJ...546..576N}, \citet{2001ApJ...546..585S} and \citet{2019ApJ...880...13Z} for the evaluation of mode excitation rates. For all target stars, components of the expression of excitation rate are computed directly from the corresponding 3D model atmosphere and patched 1D stellar structural model.
The expression of linear damping rate was derived from the first-order perturbation theory where non-adiabatic effects are treated as small perturbations to adiabatic oscillations. In order to extract reliable damping rates from 3D simulations using analytical formula \eqref{eq:etamain}, it is necessary to separate the density fluctuation $\delta\rho$ caused by pulsation from ``convective noise.'' To this end, we carried out artificial mode driving simulations where a target radial oscillation is artificially driven to large amplitude so that $\delta\rho$ at the driving frequency results predominantly from pulsation. Theoretical damping rates at each frequency are computed from such numerical experiments and compared in detail with observed frequency-dependent line widths. Encouraging agreement is achieved between simulations and observation: For all four stars investigated, our numerical damping rates are of the same order of magnitude as the corresponding measured values, and in most cases the $\eta - \nu$ relation matches observation. This validates our numerical approach for calculating the damping rate for radial modes.
 
  Based on the excitation and damping rates, we calculated the velocity amplitudes from which theoretical $\nu_{\max}$ values are estimated. Our results are compared against the corresponding observations. In particular, the estimated $\nu_{\max}$ is consistent overall with measured values. This finding foreshadows exciting opportunities for predicting this important asteroseismic observable for solar-type oscillators from first principles using 3D hydrodynamical simulations.

  Studying several stars also allows for comparison between excitation/damping rates and fundamental stellar parameters. Qualitative relationships between $\mathcal{P}_{\rm exc}$, $\eta$, $\nu_{\max}$ and $T_{\rm eff}$, $g$ were found from our simulations. Namely, both excitation and damping rates are positively correlated with the effective temperature, which accords with empirical relations derived from other theoretical investigations \citep{2007A&A...463..297S} and those summarised from observations \citep{2009A&A...500L..21C}. Meanwhile, theoretical $\nu_{\max}$ values estimated from our simulations broadly obey the $\nu_{\max}$ scaling relation, reaffirming it at solar-metallicity.
  
  These facts, in combination with the results for the Sun \citep{2019ApJ...880...13Z}, suggest that our method of modelling the excitation and damping of solar-like radial mode oscillations is valid across a wide range of effective temperatures and surface gravities, especially given that there are no tunable free parameters in our formulations used to ``fit'' the theoretical results to observational data. In addition, our method enables deeper understanding of the underlying physics behind mode excitation and damping. The former was discussed in detail in \citet{2019ApJ...880...13Z}. In this work, we explored where exactly radial oscillations are damped in the near-surface region based on the artificial mode driving simulations and concluded that intermediate- and high-frequency modes are mostly damped just below the photosphere, whereas low-frequency modes demonstrate broader regions of growth and damping. The physics of the damping rate dip near $\nu_{\max}$ was also discussed, and we have addressed the question of why the damping rate dip becomes less pronounced in warmer stars -- thermal processes, which tend to destabilize modes and cause the dip near $\nu_{\max}$, have relatively lesser impact on the total damping rate in warmer stars.
  
  We caution however, that disagreements do exist between simulations and observations. For example, for Procyon, it seems that 3D simulations overestimate excitation rates. At lower frequencies, we tend to underestimate damping rates because of the limited simulation domain. All these indicators suggest room for improvement to our numerical methods or indicate the necessity of more detailed numerical simulations. More detailed 3D surface (magneto-)convection simulations with higher numerical resolution and deeper vertical coverage may be helpful towards improving the agreement between theory and observation. Meanwhile, a larger number of such simulations that covers adequate $\{ T_{\rm eff}, \log g, \rm [Fe/H] \}$ combinations could be used to quantify the relationship between mode excitation/damping and fundamental stellar parameters. It may even be possible to quantify the departure, if any, from the widely-used $\nu_{\max}$ scaling relation from an entirely theoretical angle.

\section*{Acknowledgements}

  The authors are grateful to Yaguang Li and Earl Bellinger for providing observational data, and to Thomas Nordlander, J{\o}rgen Christensen-Dalsgaard and Tim Bedding for reading and commenting on this manuscript. Anish Amarsi's help on the \textsc{blue} opacity package is greatly appreciated. We thank also Christoph Federrath, Luca Casagrande, Tim White, Aaron Dotter, and G{\"u}nter Houdek for valuable comments and fruitful discussions. M.A. gratefully acknowledges funding from the Australian Research Council (grant DP150100250). Funding for the Stellar Astrophysics Centre is provided by The Danish National Research Foundation (grant agreement no.:DNRF106). This research was undertaken with the assistance of resources provided at the NCI National Facility systems at the Australian National University through the National Computational Merit Allocation Scheme supported by the Australian Government.




\bibliographystyle{mnras}
\bibliography{References} 



\appendix
\section{Linear damping rates} \label{sec:eta}

In this appendix, we derive the expression for the linear damping rate used in the main text from basic fluid equations, and elaborate on the numerical technique we developed to extract the damping rates from 3D simulations.

\subsection{Theoretical formulation} \label{sec:etath}

  It is necessary to investigate how non-adiabatic processes will affect stellar oscillation for the calculation of the damping rate. In this section, we employ a simplified approach to include non-adiabatic effects on radial p-mode by regarding them as small perturbations. The perturbation will shift the p-mode frequency and introduce an exponential attenuation term in the mode amplitude. The latter is relevant to the damping rate. Similar discussions and derivations can be found also in, e.g., \cite{1983bhwd.book.....S} and \cite{2010aste.book.....A}.
  
  We begin with the fluid momentum equation, and, assuming that the system is subjected to no external force other than gravity,
\begin{equation}
\frac{dv^j}{dt} = -\frac{1}{\rho} \nabla_j P - \nabla_j \Phi,
\end{equation}
where $v$, $\rho$, $P$ and $\Phi$ are fluid velocity, density, pressure and gravitational potential, respectively. The index $j$ denotes the three components in Cartesian coordinates. The perturbed momentum equation then writes
\begin{equation} \label{eq:osc}
\delta \left( \frac{dv^j}{dt} + \frac{1}{\rho} \nabla_j P + \nabla_j \Phi \right) = 0,
\end{equation}
with the symbol $\delta$ representing Lagrangian perturbation (Eulerian perturbation is denoted by superscript $^{\prime}$). Expanding Eq.~\eqref{eq:osc} gives
\begin{equation} \label{eq:oscexpan}
\begin{aligned}
& \frac{d^2 \xi^j}{dt^2} 
- \frac{\delta \rho}{\rho_0^2} \nabla_j P_0 + \frac{1}{\rho_0}\nabla_j (\delta P) - \frac{1}{\rho_0}(\nabla_j \xi^k)(\nabla_k P_0)
\\
& + \nabla_j \Phi^{\prime} + \nabla_j \xi^k \nabla_k \Phi_0 - (\nabla_j \xi^k)(\nabla_k \Phi_0) = 0.
\end{aligned}
\end{equation}
Here $\Vec{\xi}$ is the fluid displacement vector, and subscript ``0'' stands for quantities in equilibrium state. The relation between Eulerian and Lagrangian perturbation is used to obtain Eq.~\eqref{eq:oscexpan}. Also, the Einstein summation convention is applied throughout this section unless otherwise specified. Eq.~\eqref{eq:oscexpan} can be simplified using the hydrostatic equilibrium equation to
\begin{equation} \label{eq:hydros}
\nabla_j P_0 + \rho_0 \nabla_j \Phi_0 = 0.
\end{equation}
The perturbation to density is related to the fluid displacement vector $\vec{\xi}$ by the perturbed fluid continuity equation
\begin{equation} \label{eq:osccont}
\delta \rho = -\rho_0 \nabla_k\xi^k.
\end{equation}
And for radial perturbations of a spherical star, the relation between perturbed gravitational potential and fluid displacement is (\citealt{1983bhwd.book.....S} chapter 6.3)
\begin{equation} \label{eq:oscPois}
\nabla_j \Phi^{\prime} = -4\pi G \rho_0 \xi^j,
\end{equation}
where $G$ is the gravitational constant. Plugging Eqs.~\eqref{eq:hydros}, \eqref{eq:osccont} and \eqref{eq:oscPois} into Eq.~\eqref{eq:oscexpan}, we have 
\begin{equation} \label{eq:chraceq}
\begin{aligned}
&\frac{d^2 \xi^j}{dt^2}
+ \frac{1}{\rho_0} (\nabla_k\xi^k) \nabla_j P_0 
+ \frac{1}{\rho_0} \nabla_j (\delta P)
\\
&- 4\pi G \rho_0 \xi^j + \nabla_j \xi^k \nabla_k \Phi_0 = 0.
\end{aligned}
\end{equation}
In the case of adiabatic oscillation, the pressure fluctuation $\delta P$ is connected with fluid displacement via
\begin{equation} \label{eq:deltaPad}
\frac{\delta P_{\rm ad}}{P_0} = \Gamma_{1,0}\frac{\delta\rho}{\rho_0}
= -\Gamma_{1,0}\nabla_k \xi^k,
\end{equation}
where $\Gamma_1 = (\partial\ln P / \partial\ln \rho)_{\rm ad}$ is the (first) adiabatic index with subscript ``ad'' representing fixed entropy. From the relation \eqref{eq:deltaPad} one can recognise that Eq.~\eqref{eq:chraceq} is the characteristic equation of the eigenvalue problem in the scenario of adiabatic oscillation. However, when considering non-adiabatic oscillations, the expression of $\delta P$, based on the perturbed energy equation (\citealt{2010aste.book.....A} Eq.~3.47), becomes
\begin{equation} \label{eq:energyeq}
\begin{aligned}
\partial_t \delta P &= \frac{\Gamma_{1,0} P_0}{\rho_0}\partial_t \delta\rho + \rho_0 (\Gamma_{3,0} - 1)\partial_t \delta q 
\\
&= \partial_t \delta P_{\rm ad} + \rho_0 (\Gamma_{3,0} - 1)\partial_t \delta q,
\end{aligned}
\end{equation}
with $q$ being heating or cooling and $\Gamma_{3,0} - 1 = (\partial\ln T / \partial\ln\rho)_{\rm ad}$. The second term on the right-hand side of Eq.~\eqref{eq:energyeq} represents the time derivative of the pressure fluctuation associated with non-adiabatic effects, that is,
\begin{equation} \label{eq:deltaP}
\partial_t \delta P = \partial_t \delta P_{\rm ad} + \partial_t \delta P_{\rm nad}.
\end{equation}
If we regard non-adiabatic effects as small perturbation, it is reasonable to assume that the time dependence of $\xi$, $\delta P$ and $\delta P_{\rm ad}$ have the form $\exp (i\omega t)$ ($\omega$ and $t$ are angular frequency and time, respectively), and Eqs.~\eqref{eq:chraceq} and \eqref{eq:deltaP} simplify into\footnote{Assuming time dependence $\exp (i\omega t)$ or $\exp (-i\omega t)$ has no physical significance, it will not affect the final damping rate expression. Also worth noting is that $\delta P_{\rm nad}$ stems from non-adiabatic effects including entropy fluctuation and convective turbulence, which are stochastic rather than coherent \citep{2001ApJ...546..585S,2019ApJ...880...13Z}. Therefore the temporal dependence of $\delta P_{\rm nad}$ is not $\exp (i\omega t)$.}
\begin{align}
\begin{split}
\omega^2 \xi^j &=
\frac{1}{\rho_0} (\nabla_k\xi^k) \nabla_j P_0 + \frac{1}{\rho_0} \nabla_j (\delta P)
\\
& - 4\pi G \rho_0 \xi^j + \nabla_j \xi^k \nabla_k \Phi_0,
\end{split}
\label{eq:chraceqfin}
\\
\delta P &= \delta P_{\rm ad} + \frac{1}{i\omega} \partial_t \delta P_{\rm nad}.
\label{eq:deltaPfin}
\end{align}
Further assume that non-adiabatic pressure fluctuation responds linearly to fluid displacement; then Eqs.~\eqref{eq:chraceqfin} and \eqref{eq:deltaPfin} can be written in the form
\begin{equation} \label{eq:eigeneq}
\mathcal{H} \ket{\xi} 
= \left[ \mathcal{H}^{(0)} + \mathcal{H}^{(1)} \right] \ket{\xi}
= \omega^2 \ket{\xi},
\end{equation}
with
\begin{align}
\begin{split}
\mathcal{H}_{jk}^{(0)} \xi^k &=
\frac{1}{\rho_0} (\nabla_k\xi^k) \nabla_j P_0 
- \frac{1}{\rho_0} \nabla_j (\Gamma_{1,0}P_0 \nabla_k\xi^k)
\\
& -4\pi G \rho_0 \xi^j + \nabla_j \xi^k \nabla_k \Phi_0, 
\label{eq:H0}
\end{split}
\\ 
\mathcal{H}_{jk}^{(1)} \xi^k &=
\frac{1}{\rho_0} \nabla_j 
\left( \frac{1}{i\omega} \partial_t \delta P_{\rm nad} \right).
\label{eq:H1}
\end{align}
Here we followed the notation from quantum mechanics: $\ket{\xi}$ denotes eigenfunction, $\mathcal{H}^{(0)}$ is the operator (acting on $\xi$) in the case of adiabatic oscillation, whereas $\mathcal{H}^{(1)}$ accounts for non-adiabatic effects. The eigenfunctions form an orthogonal basis in Hilbert space \citep{1979ApJ...232..874S}, with the inner product defined as (\citealt{2010aste.book.....A} Eq.~3.246)
\begin{equation} \label{eq:innerp}
\braket{\xi_a}{\xi_b} \equiv \int_V \: \rho_0 \xi_{a,k}^*\xi_b^k \: dV,
\end{equation}
where ``$a$'' and ``$b$'' label a specific normal mode and $\xi_a^*$ is the complex conjugate of $\xi_a$, and $V$ is the volume of star.

  Now we solve Eq.~\eqref{eq:eigeneq} with the perturbation theory (\citealt{2010aste.book.....A} chapter 3.6), and focus on a specific radial p-mode:
\begin{equation}
\mathcal{H}^{(0)} \ket{\xi_p} = \omega_p^2 \ket{\xi_p},
\end{equation}
where $\omega_p$ and $\ket{\xi_p}$ are the corresponding mode adiabatic eigenfrequency and adiabatic eigenfunction of this p-mode. The non-adiabatic term at this frequency writes
\begin{equation} \label{eq:H1fin}
\mathcal{H}^{(1)} \ket{\xi_p} 
= \frac{1}{\rho_0} \frac{1}{i\omega_p} 
\left. \mathcal{F}\left( \nabla_j \partial_t \delta P_{\rm nad} \right) \right\vert_{\omega = \omega_p} 
= \frac{1}{\rho_0} \nabla_j \delta P_{\rm nad}(\omega_p).
\end{equation}
The first-order frequency shift due to non-adiabatic effect, as given by the perturbation analysis, is then
\begin{equation}
\omega^2 - \omega_p^2 
= \frac{\bra{\xi_p} \mathcal{H}^{(1)} \ket{\xi_p}}{\braket{\xi_p}{\xi_p}}.
\end{equation}
Substituting Eq.~\eqref{eq:H1fin} and the inner product \eqref{eq:innerp} into the equation above, we get
\begin{equation} \label{eq:fresft}
\omega^2 - \omega_p^2 
= \frac{\int_V \: \xi_{p}^{*k} \nabla_k (\delta P_{\rm nad}) \: dV}
{\int_V \: \rho_0 \xi_{p}^{*k}\xi_{p,k} \: dV}.
\end{equation}
On the left-hand side, the perturbed frequency $\omega$ contains a real part and an imaginary part, that is, $\omega = \omega_{\rm Re} + i\omega_{\rm Im}$. Therefore, the perturbed radial p-mode eigenfunction has the form $\xi = \tilde{\xi} \exp (i\omega t) = \tilde{\xi} \exp (i\omega_{\rm Re} t) \exp(-\omega_{\rm Im}t)$, where the exponential part governs the change of mode amplitude. In the circumstance of mode damping, a positive damping rate should correspond to the decay of mode amplitude, hence $\eta = \omega_{\rm Im}$. Because $\omega_{\rm Re}$ is very close to unperturbed (adiabatic) frequency and $\eta \ll \omega_p$, Eq.~\eqref{eq:fresft} turns out to be
\begin{equation} \label{eq:etaori}
\eta \approx \mathfrak{Im} \left\{ 
\frac{\int_V \: \xi_{p}^{*k} \nabla_k (\delta P_{\rm nad}) \: dV}
{2\omega_p \int_V \: \rho_0 \xi_{p}^{*k}\xi_{p,k} \: dV} \right\}.
\end{equation}
Integrating the right-hand side of Eq.~\eqref{eq:etaori} by parts gives
\begin{equation}
\eta \approx \mathfrak{Im} \left\{
\frac{\int_V \: \nabla_k (\xi_{p}^{*k} \delta P_{\rm nad}) \: dV}
{2\omega_p \int_V \: \rho_0 \xi_{p}^{*k}\xi_{p,k} \: dV}
- \frac{\int_V \: (\nabla_k\xi_{p}^{*k}) \delta P_{\rm nad} \: dV }
{2\omega_p \int_V \: \rho_0 \xi_{p}^{*k}\xi_{p,k} \: dV} \right\}.
\end{equation}
Applying the divergence theorem to the first term and the perturbed fluid continuity equation \eqref{eq:osccont} to the second term, we get
\begin{equation}
\eta \approx \mathfrak{Im}\left\lbrace
\frac{\oint_{\rm surf} \: \vec{\xi}_p^* \delta P_{\rm nad} \: d\vec{S}}
{2\omega_p \int_V \: \rho_0 \xi_{p}^{*k}\xi_{p,k} \: dV}
+ \frac{\int_V \: (\delta\rho^* / \rho_0) \delta P_{\rm nad} \: dV }{2\omega_p \int_V \: \rho_0 \xi_{p}^{*k}\xi_{p,k} \: dV}
\right\rbrace,
\end{equation}
where the integration over the stellar surface is often neglected (cf.~\citealt{2010aste.book.....A} chapter 3.7 and \citealt{2001ApJ...546..576N} Sect.~3), therefore
\begin{equation} \label{eq:eta}
\eta \approx \frac{\int_V \: \mathfrak{Im}\left\lbrace (\delta\rho^* / \rho_0) \delta P_{\rm nad}  \right\rbrace \: dV}
{2\omega_p\int_V \: \rho_0 \xi_{p}^{*k}\xi_{p,k} \: dV},
\end{equation}
which is the full expression of the (linear) damping rate from first-order perturbation analysis. We note that Eq.~\eqref{eq:eta} is essentially equivalent to $\eta$ derived in previous investigations such as \cite{2012A&A...540L...7B} and \cite{2015LRSP...12....8H}.

  Next we rearrange and simplify Eq.~\eqref{eq:eta} into a different form that is more suitable for numerical evaluation. The denominator of \eqref{eq:eta} is proportional to the mode kinetic energy. For radial modes, it is related with mode mass per unit surface area $m_{\rm mode}$ and mode velocity amplitude at the photosphere $\mathcal{V}(R_{\rm phot})$ by \cite{2001ApJ...546..576N} Eq.~63:
\begin{equation} \label{eq:modeE}
\begin{aligned}
\int_V \: \rho_0 \xi_{p}^{*k}\xi_{p,k} \: dV
&= 4\pi R_{\rm phot}^2 \int_r \: \rho_0 |\xi_{p}|^2 \frac{r^2}{R_{\rm phot}^2} \: dr
\\
&= 4\pi R_{\rm phot}^2 \frac{2m_{\rm mode} |\mathcal{V}(R_{\rm phot})|^2}{\omega_p^2},
\end{aligned}
\end{equation}
where $R_{\rm phot}$ is the photosphere radius. On the other hand, the integral in the numerator of \eqref{eq:eta} is the so-called ``work integral'', representing energy transfer between convection and oscillations. Ideally, the work integral is finite throughout the entire star. In practice, however, the integral is confined within the simulation domain, outside which $\delta P_{\rm nad}$ and $\delta\rho$ are not available. Because the 3D simulation covers only a small part of the star near the photosphere, and in subsequent calculations we consider horizontally averaged non-adiabatic fluctuations and density fluctuations, the work integral reduces to
\begin{equation} \label{eq:workint}
4\pi R_{\rm phot}^2 \int_{y_{\rm bot}}^{y_{\rm top}} \: \mathfrak{Im}
\left\{ (\delta\bar{\rho}^{*} / \bar{\rho}_0) \delta \bar{P}_{\rm nad} \right\} dy,
\end{equation}
where the bar symbol denotes horizontal averaging, $y$ is the vertical direction in the Cartesian coordinate in which the 3D simulations are set, and $y_{\rm top}$ ($y_{\rm bot}$) is the geometric depth at the top (bottom) of the simulation domain. Meanwhile, it is worth noting that the near-surface region covered by 3D simulation is where non-adiabatic effects and local compression due to oscillation are the strongest. In the deep stellar interior that is outside the simulation domain, physical processes are very close to adiabatic, and local compression is weaker compared with the near-surface region. Consequently, omitting the work integral in the deep interior might not be a significant simplification. Substituting \eqref{eq:modeE} and \eqref{eq:workint} into Eq.~\eqref{eq:eta} gives
\begin{equation} \label{eq:etanum}
\eta \approx \frac{\omega_p
\int_{y_{\rm bot}}^{y_{\rm top}} \mathfrak{Im}\left\{ (\delta\bar{\rho}^{*} / \bar{\rho}_0) \delta \bar{P}_{\rm nad} \right\} dy }
{4 m_{\rm mode} \vert \mathcal{V}(R_{\rm phot})\vert^2},
\end{equation}
which is the equation we adopted for numerical computation.

\subsection{Numerical evaluation} \label{sec:etanum}

\begin{figure*}
\subfigure[]{
\begin{overpic}[width=0.48\textwidth]{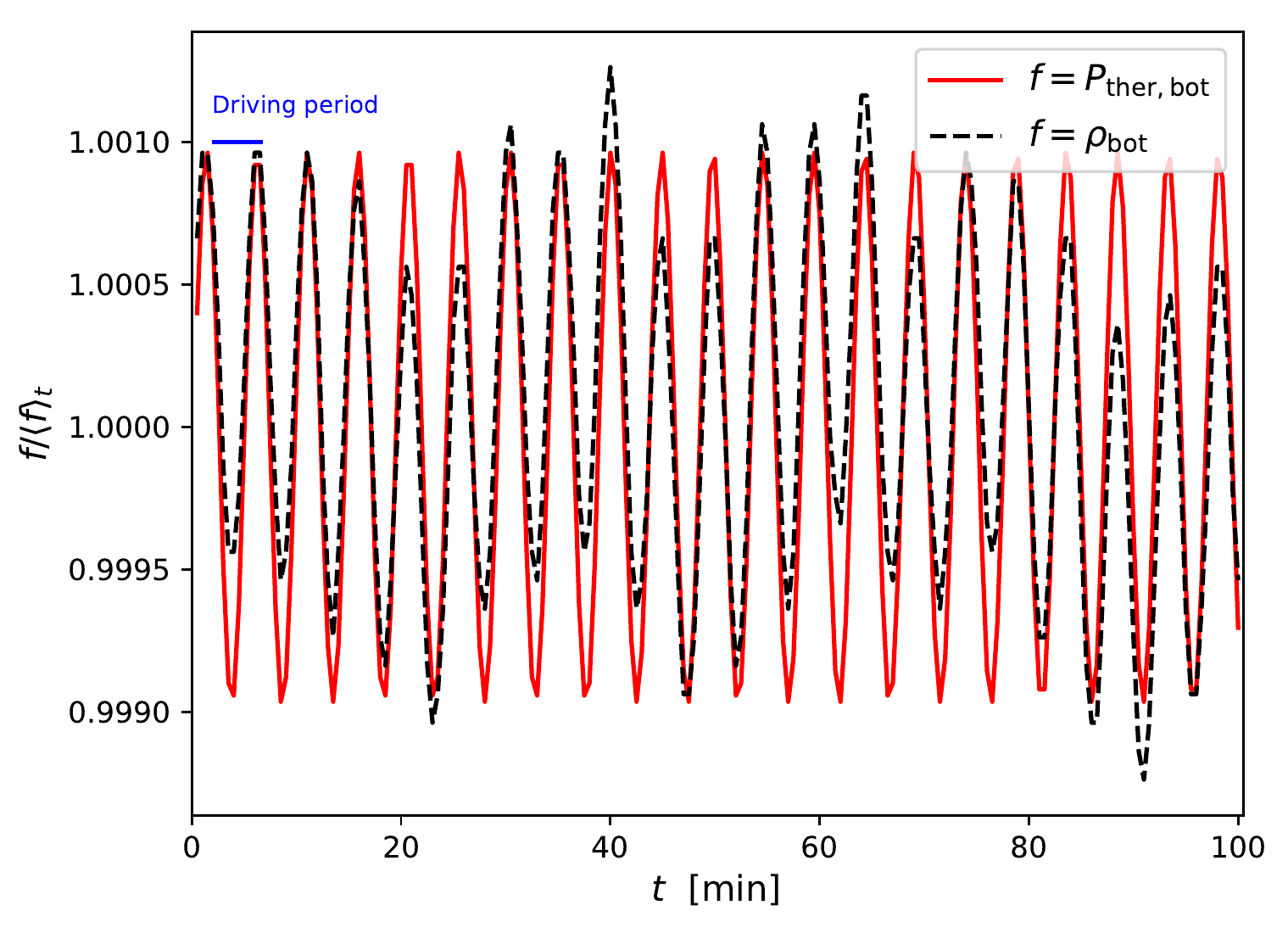}
\end{overpic}
\label{fig:sun_t_bottom}
}
\subfigure[]{
\begin{overpic}[width=0.48\textwidth]{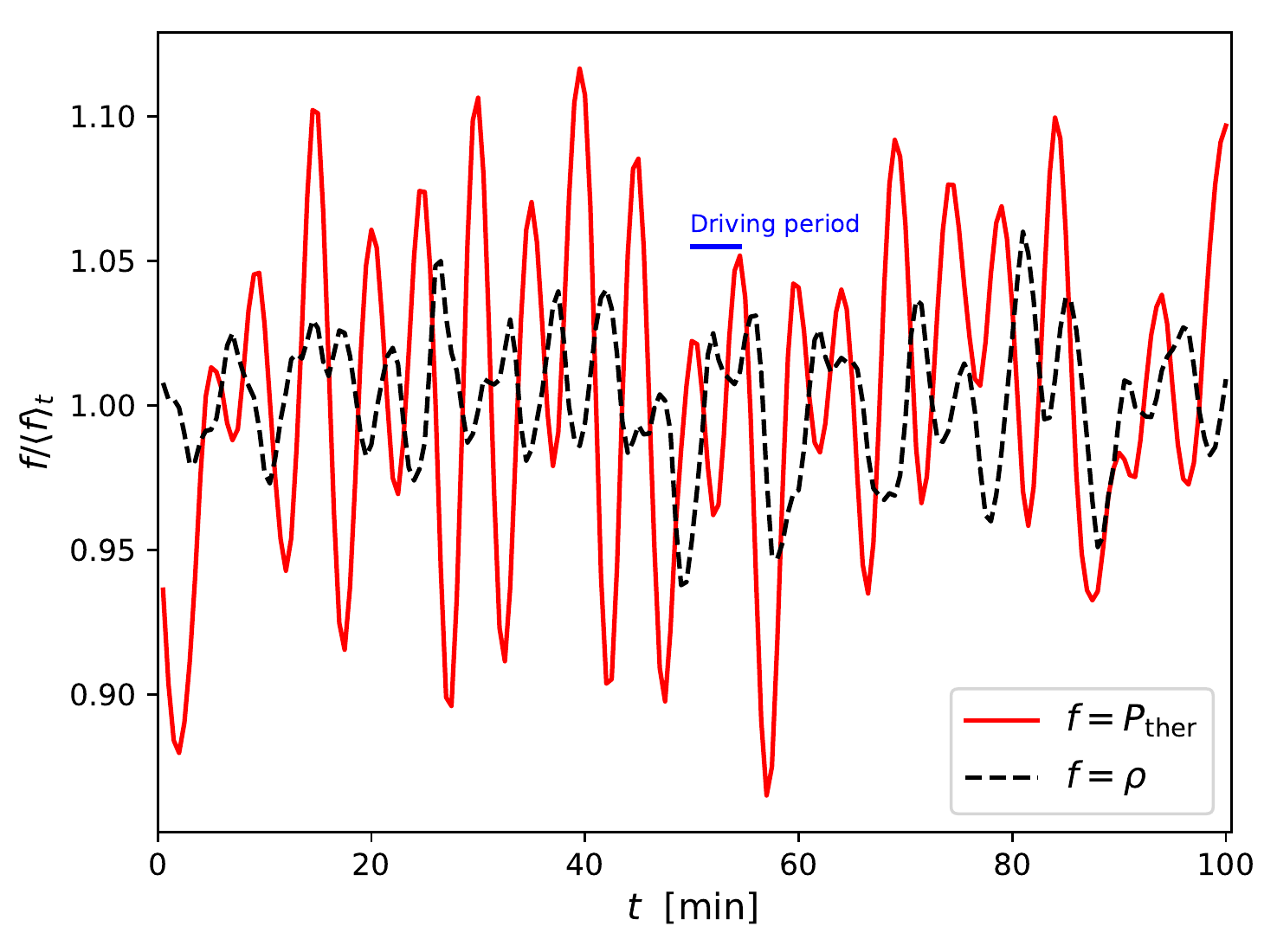}
\end{overpic}
\label{fig:sun_t_photosphere}
}
\caption{\ref{fig:sun_t_bottom}: Time variation of thermal pressure (red solid line) and density (black dashed line) at the bottom boundary of a solar 3D simulation with artificial mode driving. The input perturbation amplitude of this simulation is set to $\epsilon = 0.001$, while the period of artificial driving, which corresponds to a cyclic frequency $\nu_{\rm d} = 3.44$ mHz, is marked in the figure with blue ruler. Both the magnitude and the period of fluctuations are consistent with input values. $\langle ... \rangle_t$ stands for time averaging. \ref{fig:sun_t_photosphere}: Similar to \ref{fig:sun_t_bottom}, but showing time variation of thermal pressure and density near the photosphere.}
\end{figure*}

\begin{figure*}
\subfigure[]{
\begin{overpic}[width=0.48\textwidth]{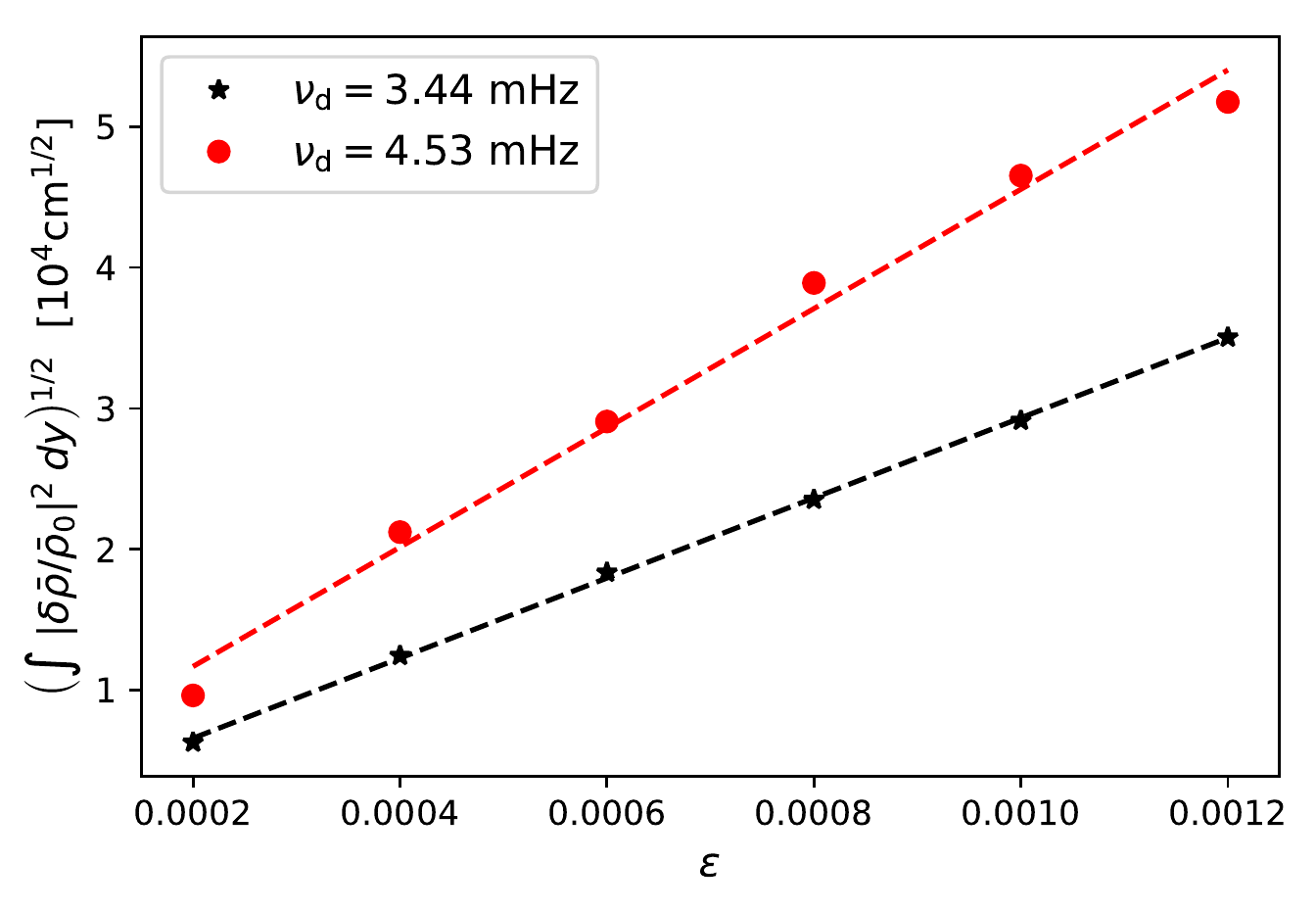}
\end{overpic}
\label{fig:deltarho_amp}
}
\subfigure[]{
\begin{overpic}[width=0.48\textwidth]{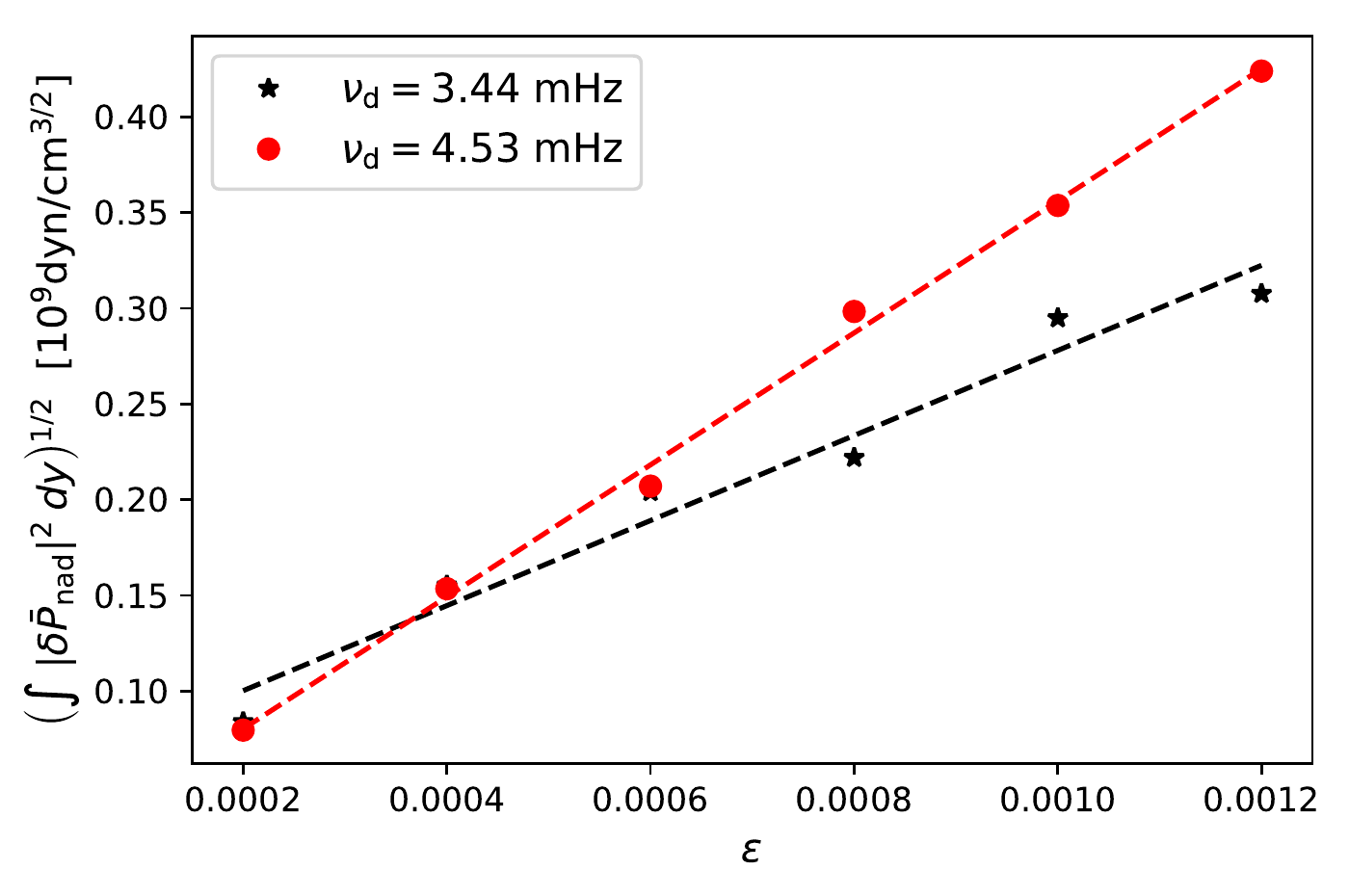}
\end{overpic}
\label{fig:deltaP_amp}
}
\caption{\ref{fig:deltarho_amp}: Integrated density fluctuations at driving frequency for different perturbation amplitudes $\epsilon$. The power spectrum of $\delta\bar{\rho}/\bar{\rho}_0$ is integrated over vertical direction of the entire simulation domain then we take the square root. Black asterisks and red dots are calculated from solar 3D simulations with different driving cyclic frequencies $\nu_{\rm d}$. Linear fits to these data points are presented in dashed lines. \ref{fig:deltaP_amp}: Similar to \ref{fig:deltarho_amp}, but showing integrated non-adiabatic pressure fluctuations at driving frequency for different perturbation amplitudes.}
\end{figure*}

  In this section we describe the numerical methods applied to compute the linear damping rates from 3D simulations. Based on Eq.~\eqref{eq:etanum}, four components -- density fluctuation $\delta\bar{\rho}$, non-adiabatic pressure fluctuation $\delta \bar{P}_{\rm nad}$, mode mass per unit surface area $m_{\rm mode}$ and velocity amplitude at photosphere $\mathcal{V}(R_{\rm phot})$ -- are essential in order to calculate $\eta$. Among them, the coupling between density and non-adiabatic pressure fluctuation represents the interaction between oscillation and convection, with the former reflecting fluid compression results from mode displacement (Eq.~\eqref{eq:osccont}) while the latter mainly stems from convective turbulence. Although both $\delta\bar{\rho}$ and $\delta \bar{P}_{\rm nad}$ are available from the 3D model, owing to the complexity of physical processes occurring in the simulation domain, $\delta\bar{\rho}$ computed from the simulation comprises not only pulsation signals, but also the signature of ``convective noise.'' To obtain an $\delta\bar{\rho}$ that cleanly reflects the contribution from the mode eigenfunction -- in other words, a coherent density fluctuation -- we conduct numerical experiments that artificially drive radial mode at a particular frequency to large amplitude. The target mode will thus be prominent in the simulation box and distinguishable from ``convective noise.''

The artificial driving is achieved by modifying the bottom boundary condition of the simulation. We adopt open bottom boundary conditions in our simulation, where the outgoing flow is free to carry entropy fluctuations out of the simulation domain. In ``normal'' simulations, incoming flows are forced to have fixed entropy and thermal pressure, which are constant over the horizontal plane. However, in the artificial driving experiment, we impose a small time-dependent perturbation to the thermal pressure at the bottom boundary so that the pressure of incoming flows fluctuates with time. Meanwhile, we enforce constant entropy (to first order) of the incoming flows at the bottom boundary to ensure no extra energy is injected into the system. Namely,
\begin{equation} \label{eq:bbc}
\begin{aligned}
P_{\rm bot} &= P_{\rm bot,0} \left[1 + \epsilon\sin(\omega_{\rm d} t + \phi)\right],
\\
s_{\rm bot} &= s_{\rm bot,0} + \mathcal{O}(\epsilon^2),
\end{aligned} 
\end{equation} 
where $P_{\rm bot,0}$ and $s_{\rm bot,0}$ are constant thermal pressure and constant entropy (per mass) at the bottom boundary of ``normal'' simulations. The term $\epsilon$ is a small, dimensionless number that governs the amplitude of perturbation, $\omega_{\rm d}$ is the angular frequency of artificial driving, and $\phi$ is the phase. The applied perturbation varies sinusoidally with time and remains uniform over the horizontal plane, since radial oscillations are the focus here.

According to \eqref{eq:bbc}, other thermochemical quantities, such as mass density and energy density, will also fluctuate coherently with thermal pressure at the bottom boundary (displayed in Fig.~\ref{fig:sun_t_bottom}). As a result, the perturbation will generate coherent fluid motion with the same frequency as the driving frequency $\omega_{\rm d}$, and amplify vertical velocity, density and pressure fluctuation to large magnitudes in the entire simulation domain (Fig.~\ref{fig:sun_t_photosphere}).
Given that none of $\delta\bar{\rho}$, $\delta\bar{P}_{\rm nad}$ and $\mathcal{V}(R_{\rm phot})$ are realistic from artificial driving, it is natural to question whether one could obtain reliable damping rates from such a numerical experiment. The answer to this question is affirmative if $\delta\bar{\rho}$, $\delta\bar{P}_{\rm nad}$ and $\mathcal{V}(R_{\rm phot})$ all respond linearly to mode displacement. 
In this scenario, the unrealistically large oscillation amplitude resulting from artificial driving cancels out if we calculate $\eta$ via Eq.~\eqref{eq:etanum}. Because velocity is the time derivative of mode displacement, it is linearly proportional to the mode amplitude that is controlled by $\epsilon$. The density fluctuation is related to mode displacement by \eqref{eq:osccont}, which is a linear relation as well. However, the linearity between $\delta\bar{P}_{\rm nad}$ and mode displacement -- upon which we also rely when deriving the linear damping rate from perturbation theory (see Sect.~\ref{sec:etath}) -- is not apparent. To this end, we compute $\delta\bar{P}_{\rm nad}$ for different perturbation amplitudes $\epsilon$ (all other input parameters are exactly the same except for $\epsilon$ in order to control variables), and depict $\delta\bar{P}_{\rm nad}$ at the driving frequency as a function of $\epsilon$ in Fig.~\ref{fig:deltaP_amp}. As observed from Fig.~\ref{fig:deltaP_amp}, $\delta\bar{P}_{\rm nad}$ responds nearly linear to $\epsilon$ between $\epsilon = 0.0002$ and 0.0012, suggesting an approximately linear relationship between non-adiabatic pressure fluctuation and mode displacement. Thus, we can claim that such artificial driving simulations are able to give reliable damping rate results because the rates of $\delta\bar{\rho}$, $\delta\bar{P}_{\rm nad}$ and $\mathcal{V}(R_{\rm phot})$ enhancement are similar to each other, so that the artificial effect from ``mode driving'' largely cancels out between $(\delta\bar{\rho}^{*} / \bar{\rho}_0) \delta \bar{P}_{\rm nad}$ and $\vert \mathcal{V}(R_{\rm phot}) \vert^2$ when we compute damping rate at the driving frequency using Eq.~\eqref{eq:etanum}.

  Such numerical experiments are repeated at different driving frequencies to obtain theoretical damping rates as a function of frequency. For the purpose of controlling variables, for a given star, all artificial driving experiments are carried out with the same time duration (roughly 40-50 period of the longest driving period, that is, the lowest driving frequency) and perturbation amplitude. Other input parameters are also identical except for the driving frequency. The exact values of driving frequencies are determined based on three constraints. First, because in this work we are interested in modes with frequencies close to the observed $\nu_{\max}$, the selected driving frequencies range from approximately $2\nu_{\max} - \nu_{\rm ac}$ to $\nu_{\rm ac}$, where $\nu_{\rm ac}$ is the acoustic cut-off frequency of our target star which can be estimated from the seismic scaling relation (e.g., Eq.~1 of \citealt{2011A&A...530A.142B}). Second, driving frequencies are chosen to be close to the measured $l=0$ mode frequencies of the target star. Additionally, we require that the whole simulation period be an integer multiple of the driving period so that the damping rate does not depend on the phase of the driving, and no interpolation is needed when processing the Fourier transformed simulation data. All artificial mode driving simulations are initiated from the same snapshot that was generated in a ``normal simulation.'' The exact configurations (numerical resolutions, timespan, sampling interval etc.) of artificial mode driving simulations for our target stars are presented in Sect.~\ref{sec:3D}.

We evaluate $\eta$ in the frequency domain using numerical data produced by the artificial driving simulation. Density and non-adiabatic pressure fluctuation are computed in a pseudo-Lagrangian frame that filters out the main effects of p-mode oscillations in the simulation box (\citealt{2019ApJ...880...13Z} Sect.~3.2). In pseudo-Lagrangian frame, density fluctuation
\begin{equation}
\frac{\delta\bar{\rho}(t)}{\bar{\rho}_0} 
= \frac{\bar{\rho}_{\rm L}(t) - \bar{\rho}_{\rm 0,L}}{\bar{\rho}_{\rm 0,L}},
\end{equation}
and the non-adiabatic pressure fluctuation read (\citealt{2019ApJ...880...13Z} Eq.~13)
\begin{equation}
\begin{aligned}
\delta \bar{P}_{\rm nad}(t) =&
\left[ \left( \ln\bar{P}_{\rm L}(t) - \ln\bar{P}_{0,\rm L} \right) \right.
\\ 
&- \left.\bar{\Gamma}_{1,\rm L} \left( \ln\bar{\rho}_{\rm L}(t) - \ln\bar{\rho}_{0,\rm L} \right) \right] \bar{P}_{\rm L}.
\end{aligned}
\end{equation}
Here, quantities defined in the pseudo-Lagrangian frame are marked with subscript ``L.'' We then transfer density and non-adiabatic pressure fluctuation from time to frequency domain and take the complex conjugate of $\delta\bar{\rho}$. The imaginary part of $(\delta\bar{\rho}^{*} / \bar{\rho}_0) \delta \bar{P}_{\rm nad}$ at the driving frequency is integrated from the bottom boundary, along vertical direction, to the top boundary of the simulation domain. The result is multiplied by (angular) driving frequency to finally obtain the numerator of Eq.~\eqref{eq:etanum}. On the other side, the photospheric velocity amplitude is also evaluated from the 3D simulation. First, we average vertical velocity over the horizontal plane and compute its power spectrum. The power spectrum is then multiplied by 2 in order to convert to velocity amplitude power $|\mathcal{V}|^2$. Next, the value of $|\mathcal{V}|^2$ near optical depth unity at driving frequency is exacted to represent $|\mathcal{V}(R_{\rm phot})|^2$ in Eq.~\eqref{eq:etanum}. Mode mass per unit surface area $m_{\rm mode}$, however, is calculated from the 1D patched model using \textsc{adipls}.

\subsubsection{Effects of numerical resolution}

\begin{table}
\begin{threeparttable}
\centering
\caption{Damping rates computed at representative driving frequencies with low and normal resolution simulations. All quantities are in $\mu$Hz, and no smoothing is performed here.
\label{tb:resolution}}
{\begin{tabular*}{\columnwidth}{@{\extracolsep{\fill}}ccccc}
\toprule[2pt]

  \multicolumn{2}{c}{} & Low & Intermediate & High 
  \\
\midrule[1pt]
  \multirow{3}{*}{KIC 6225718} & $\nu_{\rm d}$ & 1852 & 2391 & 2996
  \\ 
    & $\eta$ ($120^2 \times 125$) & 2.8 & 16.4 & 30.7
  \\ 
    & $\eta$ ($240^3$) & 5.6 & 16.1 & 30.8
  \\  
\midrule[1pt]
  \multirow{3}{*}{$\beta$ Hydri} & $\nu_{\rm d}$ & 774 & 1044 & 1380
  \\
    & $\eta$ ($120^2 \times 125$) & 0.8 & 6.1 & 18.4
  \\
    & $\eta$ ($240^3$) & 1.2 & 6.3 & 16.1
  \\
\midrule[1pt]
  \multirow{3}{*}{$\delta$ Eri} & $\nu_{\rm d}$ & 541 & 691 & 869
  \\
    & $\eta$ ($120^2 \times 125$) & 1.2 & 3.3 & 11.8
  \\
    & $\eta$ ($240^3$) & 1.1 & 1.4 & 11.1
  \\
\bottomrule[2pt]
\end{tabular*}}
    
\end{threeparttable}
\end{table}

  As discussed above, the artificial driving simulation is repeated at different driving frequencies to obtain $\eta$ as a function of frequency. In other words, 20--30 such simulations are needed for each target star, which is very costly in terms of computation time and power. A promising solution to this practical issue is to carry out artificial driving simulations in lower resolution ($120^2 \times 125$). However, reducing numerical resolution, or equivalently, increasing grid spacing, will impact the small scale structure of convection and the dissipation of short wavelength fluctuations through artificial diffusion (which depends on grid spacing, see \citealt{1995...Staggercodepaper} Sect.~3 and \citealt{1998ApJ...499..914S} Sect.~2) etc. The accuracy of the work integral is likely to be affected as well because of fewer grids points in the vertical direction. Therefore, the effect of numerical resolution on damping rate results should be investigated before opting for low resolution simulations.

  Here, we study this problem in detail for KIC 6225718, $\beta$ Hydri and $\delta$ Eri. For each star, we carried out artificial driving simulations at three representative driving frequencies (one far below $\nu_{\max}$ of the corresponding star, one near $\nu_{\max}$ and one far greater than $\nu_{\max}$) with both low and normal resolution ($240^3$). In order to isolate the effect of resolution, all simulations are initiated from the same snapshot, and share the same time duration, sampling interval, and physical extent of the simulation box (see table \ref{tb:3Dsetup}), with numerical resolution being the only difference. Note that Procyon is not investigated, as it is currently not feasible to carry out multiple artificial driving simulations for this star at $240^3$ resolution with available computational resources\footnote{Convection simulation of F-type stars are much more computationally expensive than cooler stars, as they have high radiative cooling rate and strong velocity field near photosphere. Both factors lead to smaller simulation timestep (relative to the typical timescale of convection) compared to G- and K-type stars.}. Nevertheless, we are aware that numerical resolution may have an impact on the theoretical damping rate of Procyon, therefore results presented in Fig.~\ref{fig:t66g40m00eta} should be interpreted with recognition of this caveat.
  
  The results of our resolution studies are shown in Table \ref{tb:resolution}. At high frequencies, damping rates demonstrate little dependence on numerical resolution for all stars investigated. A possible explanation is that high-frequency oscillations are mostly damped in a small region near the photosphere (Fig.~\ref{fig:work_mode}) where our simulations have the highest vertical resolution so that even $120^2 \times 125$ resolution simulations are able to resolve the damping of high-frequency oscillations. In the case of KIC 6225718 and $\beta$ Hydri, damping rates evaluated at intermediate driving frequencies are not sensitive to resolution. However, for $\delta$ Eri, an obvious resolution effect is observed, which suggests that artificial driving simulation with $120^2 \times 125$ resolution is not sufficient for this star. Damping rates at low frequencies also show clear dependence on numerical resolution. Nevertheless, the resolution effect here is not of great concern because the accuracy of low-frequency theoretical damping rates are primarily limited by the vertical size of simulation domain, as already stated in the main text and demonstrated further in Sect.~\ref{sec:etay}.

  In short, for KIC 6225718 and $\beta$ Hydri, it is adequate to perform the artificial driving simulation with low resolution in order to reduce computation cost. In the case of $\delta$ Eri however, artificial driving simulations with at least $240^3$ resolution are required. Given that $\delta$ Eri is quite distinguishable from the other two stars both in terms of basic stellar parameters and evolutionary stage (Sect.~\ref{sec:star}), our resolution study might imply that at least normal numerical resolution is necessary to obtain reliable damping rates for low surface gravity subgiants and red giants.

\subsubsection{Effects of vertical extent of the simulation} \label{sec:etay}

\begin{figure}
\begin{overpic}[width=\columnwidth]{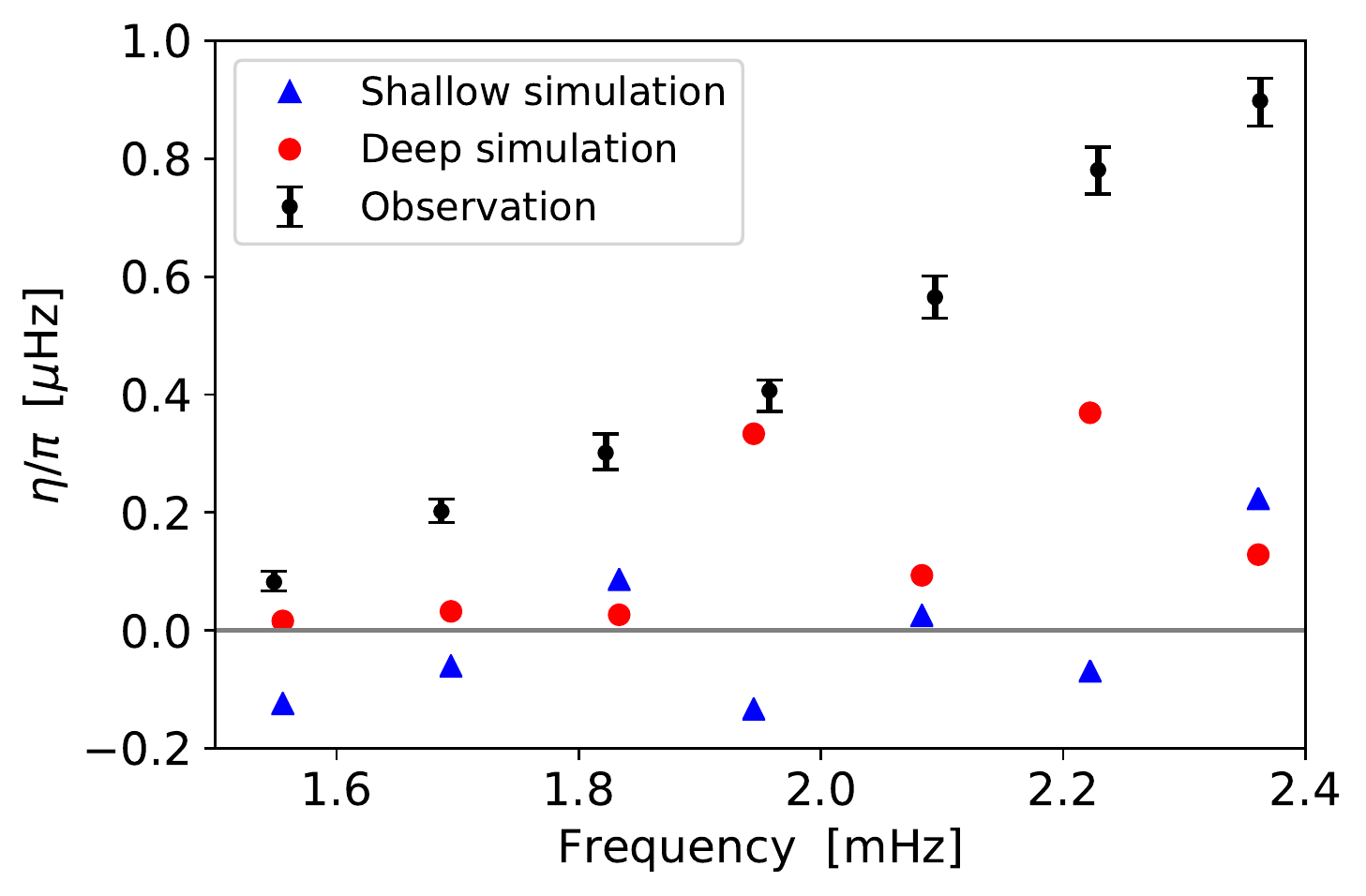}
\end{overpic}
\caption{Low-frequency damping rates computed from shallow (blue triangles) and deep (red dots) solar simulations without any smoothing. Theoretical results are divided by $\pi$ to compare with measured $l=0$ line widths from BiSON (Birmingham Solar Oscillations Network, \citealt{2005MNRAS.360..859C,2014MNRAS.439.2025D}).}
\label{fig:sun_eta_lowf}
\end{figure}

  It was stated in Sect.~\ref{sec:discuss} that damping rates at low frequencies are underestimated because of the limited size of 3D simulation. Had the 3D model been extended to deeper stellar interior, the expectation is that this discrepancy would be reduced. In this subsection, we provide substantial evidence of these assertions by comparing low-frequency theoretical damping rates computed from two sets of solar simulations with different vertical extent. The first group has the same simulation set-up as simulations used in \citet{2019ApJ...880...13Z}, which covers 3.8 Mm in the vertical direction: approximately 1 Mm above the base of the photosphere and 2.8 Mm below it. We will refer to this as the ``shallow simulation'' hereinafter. The second group, i.e.~the ``deep simulation,'' extends from approximately 0.9 Mm above the base of the photosphere to 7 Mm below. For both sets of simulations, we calculate $\eta$ at identical driving frequencies between 1.5 mHz and 2.5 mHz following the method described above. The shallow and deep simulations also share the same sampling interval (30 seconds) and total timespan (10 hours) in order to make them comparable. Low-frequency damping rates computed from two groups of simulation, together with measured solar radial mode line widths in this frequency range \citep{2005MNRAS.360..859C,2014MNRAS.439.2025D}, are presented in Fig.~\ref{fig:sun_eta_lowf}. It is obvious that damping rates computed from deep simulations are overall larger than results from shallow ones, confirming our assertions in the main text. Still, at two driving frequencies, the shallow simulations predict higher $\eta$ than the deep ones, which might be the consequence of the stochastic nature of mode damping. Also apparent is that the shallow simulation gives negative damping rates at four selected frequencies; this is in conflict with the consensus that solar radial modes are stable \citep{2015LRSP...12....8H}. This contradiction suggests low-frequency damping rates computed based on shallow simulations are not reliable, even qualitatively. This problem, however, is never seen in deep simulations.
  
  By extending the simulation domain deeper into stellar interior, the discrepancy in $\eta$ between simulation and observation is indeed reduced. Nevertheless, some uncertainty remains: damping rates predicted by deep simulations are still systematically lower than observed values for reasons that are currently unclear. The remaining uncertainty in the low frequency range will be investigated further in future work.


\bsp	
\label{lastpage}
\end{document}